\documentclass[times,twocolumn,final]{elsarticle}

\usepackage{cnf}

\usepackage{etoolbox}
\newbool{usecolors}
\setbool{usecolors}{true} 

\usepackage{framed,multirow}

\usepackage{amssymb}
\usepackage{latexsym}
\usepackage{amsmath}

\usepackage{url}
\usepackage[dvipsnames]{xcolor}
\usepackage{graphicx}
\definecolor{newcolor}{rgb}{.8,.349,.1}

\usepackage{hyperref}
\usepackage{cleveref}
\usepackage{makecell}

\usepackage[switch,pagewise]{lineno} 

\usepackage{chemmacros}
\usepackage{chemfig}
\usepackage{chemformula}
\usepackage{enumitem}

\newcommand{\mrm}[1]{\mathrm{#1}}
\newcommand{\St}{\mathrm{St}}

\newcommand{\Rel}{\mathrm{Re}_\lambda}
\newcommand{\YO}{Y_\mathrm{O_2}}
\newcommand{\taub}{\tau_\mathrm{B}}

\ifbool{usecolors}{
    \newcommand{\red}{\color{red}}
    \newcommand{\blue}{\color{blue}}

}{
    \newcommand{\red}{}
    \newcommand{\blue}{}

}

\newcommand{\upd}{\mathrm{d}}                           
\newcommand{\der}[2]{\frac{\upd #1}{\upd #2}}           

\journal{Combustion \& Flame}

\begin{document}

\verso{Hemamalini et al.}

\begin{frontmatter}

\title{Effects of preferential concentration on the combustion of iron particles – A numerical study with homogeneous
isotropic turbulence}%

\author[1,2]{Shyam \snm{Hemamalini}}
    
\author[1]{B\'en\'edicte \snm{Cuenot}}  
\author[1,2]{XiaoCheng \snm{Mi}\corref{cor1}}
\cortext[cor1]{Corresponding author}
\emailauthor{x.c.mi@tue.nl}{XiaoCheng Mi}

\address[1]{Department of Mechanical Engineering, Eindhoven University of Technology, P.O. Box 513, 5600MB, Eindhoven, The Netherlands}
\address[2]{Eindhoven Institute of Renewable Energy Systems, P.O. Box 513, 5600MB, Eindhoven, The Netherlands}

\begin{abstract}

\noindent Iron particles, with their non-volatile combustion mode, remain in the dispersed phase throughout the combustion process, causing the flow in a typical iron powder combustor to be particle-laden and turbulent. Preferential concentration is a phenomenon prevalent in such turbulent flows that causes particle clustering and may be detrimental to the combustion process. To estimate the effects of clustering on the combustion process -- the most significant of which is the extension of the combustion time -- direct--numerical-simulations (DNS) are performed on a cubical domain with forced homogeneous isotropic turbulence. Three sets of simulations pertaining to Kolmogorov Stokes number $\mathrm{St}=\left( 1,10,50 \right) $, turbulent Reynolds number $\mathrm{Re_\lambda}= \left(5,10,20\right)$, and global equivalence ratio (considering \ch{FeO} as the oxidation product) $\phi=\left(0.25,0.5,0.75\right)$ are executed. The prevalence of clustering was found to be strongly sensitive to $\mathrm{St}$, as reported in the literature. Increasing $\mathrm{Re_\lambda}$ enhances the magnitude of clustering but retains the timescales of cluster formation. Increasing $\phi$ significantly extends the completion time of combustion owing to the depletion of $\mathrm{O_2}$ in particle-rich regions. The particle combustion times are estimated for the combustion of a fully clustered distribution, a Poisson (random) distribution, and a particle-gas coupled 0D suspension model, and are compared. The Poisson distribution of particles burns faster with a higher peak mean temperature, possibly due to collective heating effects. The evolution of the mean temperature in the combustion of the clustered distribution is smooth and results in a smaller peak value. However, the total combustion time of a clustered distribution is significantly extended, by up to eight times at $\mathrm{Re_\lambda}=20$ and $\phi=0.75$. Analysis of the Vorono\"i volumes $V_\mathrm{norm}$ at the start of combustion shows that particles in highly dense regions burn longer, as seen before in the literature. Furthermore, the combustion time exhibits a strong exponential dependence on $V_\mathrm{norm}$ in the ``cluster'' regions, and an asymptotic behavior in the ``void'' regions. However, significant spread is observed in the correlation. Time-averaging $V_\mathrm{norm}$ does not minimize this variation considerably. Analysis of the macroscale $\mrm{O_2}$ depletion zone indicates the importance of the macrostructure--proximity of multiple clusters--on the extension of the combustion time.
\end{abstract}


\end{frontmatter}


\section*{Novelty and significance statement}
This study caters to the research and development of iron power technology, a novel carbon-free renewable energy storage alternative to fossil fuels. The focus of this article is the effects of particle clustering, through the phenomenon of preferential concentration, on the combustion process of iron particles. The statistical analysis methods developed in this work serve as a common framework for future computational investigations on turbulent iron-powder combustion. The in-depth quantitative analysis is novel and provides deep insight into the factors influencing the combustion of such distributions.

\section{Introduction}
\label{sec1}

Iron powders with micron-sized particles are one such novel carbon-free renewable energy carrier that has seen rapid development. At the time of writing, iron combustors capable of power generation in the range of 100 kW to 1 MW are in operation \cite{metalot2025ironpower}. Typical iron combustors on an industrial scale employ turbulent flow to improve residence time and promote the homogenization of fuel and oxidizer \cite{niekvanrooijthesis,van20260}. Unlike many other solid fuels (e.g., coal, biomass, aluminum powder), the combustion temperature of iron powder in air is lower than the boiling points of both iron and its oxides. As a result, the iron particles remain in the condensed phase throughout the combustion process, and the flow in an iron-powder combustor is characterized by a turbulent, particle-laden regime. Recent years have seen considerable research into understanding the underlying physics of single iron particle ignition \cite{mi2025theoretical,st2026iron} and combustion \cite{Fujinawa2023,Thijs2022,Ning2021,Mich2023}.

Preferential concentration is a fluid dynamic phenomenon unique to such turbulent particle-laden flows, where the particles form regions of highly dense (clusters) and highly sparse (voids) particle concentrations. This phenomenon has been studied extensively in fluid dynamics through the past and present century \cite{Squires1991,Eaton1994,Coleman2009, Monchaux2012,Sumbekova}. The occurrence of this phenomenon is related to the Stokes number $\St$, which is the ratio between particle relaxation $\tau_\mrm{p}$ and flow timescales. Classically, the flow timescale of significance is the Kolmogorov timescale $\tau_\mrm{\eta}$ \cite{Eaton1994} with particle clustering explained as a consequence of centrifugation of the particles by the turbulent flow field. However, Coleman and Vassilicos \cite{Coleman2009} propose a sweep-stick mechanism by which clustering is possible at any scale, irrespective of the Kolmogorov Stokes number $\mrm{St}$. Sumbekova \emph{et al.} \cite{Sumbekova} note that the sweep-stick mechanism is apparent only at larger turbulent Reynolds numbers where eddies at various scales are stable, and at smaller $\Rel$, the dependence of particle clustering on $\St$ is still strong. Furthermore, particle clustering at higher $\Rel$ can occur as long as the particle relaxation timescale $\tau_\mrm{p}$ is between the Kolmogorov timescale $\tau_\mrm{\eta}$ and the integral timescale $\tau_\mrm{L}$.

Preferential concentration has the potential to cause significant effects in the combustion process of iron particles, such as incomplete combustion and temperature spikes. Hence, it is of industrial importance to understand this phenomenon and its consequences. Previously, Hemamalini \emph{et al.} \cite{Hemamalini2024} and Luu \emph{et al.}\cite{Luu2024} studied the turbulent combustion of iron particles in a mixing layer and observed particle clustering. A mixing layer consists of opposing cold and hot streams of gas, with the iron particles seeded through the cold stream and ignited by the hot stream. This flow scenario, although representative of realistic designs, is not an ideal setting to study preferential concentration due to its unsteady turbulence properties. Th\"ater \emph{et al.} \cite{Thaeter2024} used turbulence forcing to generate homogeneous isotropic turbulence (HIT) to study the ignition of iron particles in such preferentially concentrated clusters. Recently, Th\"ater \emph{et al.} \cite{thater2026interaction} studied the intertwined effects of iron combustion on particle clustering and the alteration of the particle clusters through the flow induced by particle combustion. The study discovered that the overall combustion times were significantly extended at higher equivalence ratios and determined that $\mrm{O_2}$ depletion is the major cause of the extension in combustion times.

A result that has been reported by Hemamalini \emph{et al.} and Th\"ater \emph{et al.} \cite{Hemamalini2024,Thaeter2024} is a comparison of a spatial property--mean minimum spacing $\bar{\delta}_\mathrm{min}$ \cite{Hemamalini2024} or Vorono\"i volume \cite{Thaeter2024}--with a property of oxidation progress--normalized combustion time $\tau_\mrm{B}$ \cite{Hemamalini2024} or oxide mass fraction $Y_\mrm{FeO}$ \cite{Thaeter2024}. This comparison yields a scatter distribution, with both studies showing a large variation in the quantity representing iron oxidation in regions of clusters and a relatively smaller variation in void regions. However, neither study analyzed this distribution further. These comparisons are the foremost attempts to correlate spatial and combustion statistics and to analyze the effects of particle clustering on combustion.

\subsection{Problem statement\label{sec:problemstatement}}

The overarching question and research that this work adds to is: \textit{can we estimate the effects of preferential concentration on turbulent iron powder combustion? How significant are these effects?}

The following research questions (abbreviated as RQ in the article) are derived from the overarching question:
\begin{enumerate}[label=RQ:\Roman*,leftmargin=1.1cm]
\itemsep0em
    \item  How much is the combustion time $\tau_\mrm{B}$ extended for a clustered distribution compared to a random (Poisson) particle distribution? \label{item:rq2}
    \item What is the effect of the overall equivalence ratio (considering \ch{FeO} as the oxidation product) $\phi$ on the extension of the combustion time of clustered iron particles? \label{item:rq3}
    \item Can we deterministically predict the (extension of) combustion times based on the initial particle distribution? \label{item:rq4}
\end{enumerate}

For this purpose, we choose preferential concentration through the interaction of particles with Homogeneous Isotropic Turbulence (HIT) as the flow scenario. Furthermore, Vorono\"i decomposition is chosen as the method to quantify particle clustering through the metric of Vorono\"i volumes, and the combustion time of the particles is used as the metric to quantify the combustion process. To answer \ref{item:rq4}, the spatial properties of the clusters and the combustion of the particles are analyzed following the prior work \cite{Hemamalini2024,Thaeter2024}, and the following research questions are posed to simplify \ref{item:rq4}.
\begin{enumerate}[label=RQ:\Roman*,leftmargin=1.15cm,start=4]
\itemsep0em
    \item Is there an underlying trend in the correlation of the Vorono\"i volume and the particle combustion time? \label{item:rqa}
    \item What factors can improve the correlation between the Vorono\"i volume and the particle combustion time? \label{item:rqb}
    \item If spatial analysis through Vorono\"i decomposition cannot be well-correlated with the combustion time of the particle distribution, which spatial property can?\label{item:rqc}
\end{enumerate}

Although a forced HIT field is not realistic, it is a simple method to induce and maintain particle clustering in order to study its effects on the combustion process. Hence, this study is not meant to be extrapolated directly to realistic settings. However, the results of the study concerning the aforementioned research questions can still provide insight into the coupled dynamics of the two processes--the particle-laden turbulence and combustion.

\section{Methodology \& simulation setup\label{sec:methodology}}

Gas-phase point-particle direct numerical simulations (DNS) of particle-laden homogeneous isotropic turbulence (HIT) are employed in this work. Unlike other canonical flow scenarios, a forced HIT setting establishes constant turbulence properties and provides good control over the analysis of the phenomenon.

\subsection{Gas-phase setup\label{sec:gasphase}}

The gas phase is modeled as an Eulerian grid, and the compressible form of the Navier-Stokes equations is solved, including two-way coupled Lagrangian source terms of continuity, momentum, energy, and species conservation in a manner similar to Hemamalini et al. \cite{Hemamalini2024} as:
\begin{align}
    & \frac{\partial \rho}{\partial t} + \frac{\partial \rho u_j}{\partial x_j} = S_\mrm{M} \label{eq:mass} \\
    & \frac{\partial \rho u_i}{\partial t} + \frac{\partial \rho u_i u_j}{\partial x_j} = -\frac{\partial p}{\partial x_i} + \frac{\partial \tau_{ij}}{\partial x_j} + S_{\mrm{F},i} + S_{\mrm{turb},i} \label{eq:momentum} \\
    & \frac{\partial \rho e_t}{\partial t} + \frac{\partial (\rho e_t + p)u_j}{\partial x_j} = \frac{\partial (u_j \tau_{ij})}{\partial x_j} - \frac{\partial q_j}{\partial x_j} + S_\mrm{H} + S_{\mrm{turb,H}} \label{eq:energy} \\
    & \frac{\partial \rho Y_\alpha}{\partial t} + \frac{\partial \rho Y_\alpha u_j}{\partial x_j} = -\frac{\partial \rho Y_\alpha V_{\alpha j}}{\partial x_j} + S_\alpha \label{eq:species}
\end{align}
\noindent where $S_\mrm{M}$, $S_\mrm{F,i}$, $S_\mrm{H}$, and $S_\mrm{\alpha}$ are the Lagrangian source terms of density, momentum, energy, and species mass fractions, respectively. $S_{\mrm{turb},i}$ and $S_{\mrm{turb},H}$ represent the momentum and energy source terms from the HIT forcing scheme. The high-order finite-difference solver NTMIX-CHEMKIN \cite{poinsot1992boundary} is used for the simulations in this work, with eighth-order central-differencing finite-difference discretization of space and third-order Runge-Kutta discretization of time. No gas phase reactions are taken into account. 

The HIT is forced following the methodology of Eswaran and Pope \cite{Eswaran1988}. The forced turbulent field ensures statistically steady synthetic turbulence at a fixed $\Rel$ and $\eta$ in the domain. The complete procedure on how the forcing scheme is implemented is presented in Appendix A.

\subsection{Particle setup\label{sec:particle}}

The iron particles are modeled as Lagrangian point particles, similar to Hemamalini \emph{et al.} \cite{Hemamalini2024}. The particle location $\mathbf{x}_{\mathrm{p}}$ and velocity $\mathbf{u}_{\mathrm{p}}$ are tracked as:
\begin{equation}
\begin{split}
    \der{\mathbf{x}_{\mathrm{p}}}{t} &= \mathbf{u}_{\mathrm{p}}\\
    \der{\mathbf{u}_{\mathrm{p}}}{t} &= \frac{3}{4}\frac{C_\mrm{D} \rho}{d_\mathrm{p} \rho_\mathrm{p}}\left| \mathbf{u} - \mathbf{u}_{\mrm{p}} \right| \left( \mathbf{u} - \mathbf{u}_{\mathrm{p}} \right) 
\end{split}\label{eq:particledrag}
\end{equation}
The momentum exchange between the Lagrangian iron particles and the gas phase, represented by $S_{F,i}$ in Equation \ref{eq:momentum}, is determined by summing the drag forces of all particles $N_p$ within a local Eulerian grid cell with a volume $V_{\mrm{cell}}$:
\begin{equation}
    S_{F,i} = -\frac{1}{V_{\mrm{cell}}} \sum_{n=1}^{N_p} m_p \frac{d u_{p,i}}{dt} W(\mathbf{x} - \mathbf{x}_p)
\end{equation}
where $m_p$ is the particle mass and $W$ is the interpolation weight function used to distribute the point force to the Eulerian grid nodes--in this case, linear interpolation.

The reaction model is similar to the switch-type model given by Mi \textit{et al.} and Jean-Phylippe \textit{et al.} \cite{Mi2022,Jean_Philyppe_2023}, where the mass consumption of oxygen is determined as the slower of the two rate-limiting processes: solid-state diffusion of $\mathrm{Fe}$ ions through the $\mrm{FeO}$ layer and diffusion of $\mrm{O_2}$ from the bulk to the particle surface, given by Eqs. \ref{eq:mr} and \ref{eq:diffo2}.
\begin{equation}
    \dot{m}_\mrm{FeO} = \rho_\mrm{FeO} A_\mrm{p} k_\mrm{0,FeO}\frac{r_\mrm{p}- X_\mrm{FeO}}{r_\mrm{p} X_\mrm{FeO}} \exp \left( \frac{-T_\mrm{a}}{T_\mrm{p}}\right) \notag
\end{equation}
\begin{equation}
    \dot{m}_\mrm{O_2,R} = \nu_\mrm{O_2/FeO}\ \dot{m}_\mrm{FeO} \label{eq:mr}
\end{equation}
\noindent where $A_\mrm{p}$ is the particle surface area, $k_0 = 2.67\times10^{-4}\SI{}{\meter^2/\second}$ the Arrhenius pre-exponential factor, $T_\mrm{a} = \SI{20319}{\kelvin}$ the activation temperature of $\mrm{Fe}$ as given by Mi \textit{et al.} \cite{Mi2022}, $r_\mrm{p}$ the particle radius , $X_\mrm{FeO}$ the thickness of the $\mrm{FeO}$ layer, and {$\nu_\mrm{O_2/FeO} = 0.5 \mathit{MW}_\mrm{O_2}/\mathit{MW}_\mrm{FeO}\approx0.2227$ the stoichiometric mass ratio of $\mrm{O_2}$ and $\mrm{FeO}$ for the single-stage reaction.}
\begin{equation}
    \dot{m}_\mrm{O_2,D,max} = A_\mrm{p}\ \beta_\mrm{D,O_2}\ \rho_\mrm{O_2,f} \label{eq:diffo2}
\end{equation}
\noindent where $\rho_\mrm{O_2,f}$ is the density of oxygen in the film layer surrounding the particle. $\beta_\mrm{D,O_2}$ is the diffusive mass transfer coefficient and is determined as:
\begin{equation}
    \beta_\mrm{D,O_2} = \frac{\mrm{Sh}\ D_\mrm{O_2}}{d_\mrm{p}}, \quad\mrm{Sh} = 2 + 0.552\ \mrm{Re}^{0.5}\mrm{Sc}^{0.33} \label{eq:betad}
\end{equation}
A film factor of 0.5 is used to adjust $Y_\mrm{O_2}$ in the calculation of boundary layer transport rates, following Thijs et al. \cite{Thijs2022}. A "switch" model is used to select the ultimate value of $\mrm{O_2}$ consumed per time-step:
\begin{equation}
    \begin{split}
        \dot{m}_\mrm{O_2} &= \dot{m}_\mrm{O_2,R}\quad\quad\: \mrm{if}\ \dot{m}_\mrm{O_2,R}<\dot{m}_\mrm{O_2,D,max} \\
        &= \dot{m}_\mrm{O_2,D,max} \quad \mrm{otherwise} \label{eq:mo2}
    \end{split}
\end{equation}
\noindent with the volumetric value of the $\mrm{O_2}$ mass consumption rate constituting the source terms $S_\mrm{M}$ and $S_\mrm{\alpha}$ in the gas-phase equations as:
\begin{equation}
    S_M = S_\mrm{O_2} = -\frac{1}{V_{\mrm{cell}}} \sum_{n=1}^{N_p} \dot{m}_\mrm{O_2} \cdot W(\mathbf{x} - \mathbf{x}_p) 
\end{equation}

The particle enthalpy is modeled similarly to Hemamalini \emph{et al.} \cite{Hemamalini2024} as:
\begin{equation}
    \der{H_\mrm{p}}{t} = \frac{h_\mrm{O_2}}{W_\mrm{O_2}} \dot{m}_\mrm{O_2} + Q_\mrm{conv} + Q_\mrm{rad} + Q_\mrm{evap} \label{eq:enthalpysimple}
\end{equation}
where $h_\mrm{O_2}$ is the specific enthalpy of $\mrm{O_2}$ at $T_\mrm{p}$, $Q_\mrm{conv}$, $Q_\mrm{rad}$, and $Q_\mrm{evap}$ are the convective, radiative, and evaporative fluxes given by:
\begin{align}
    Q_\mrm{conv} &= A_\mrm{p} h_\mrm{p} \left( T_\mrm{g} - T_\mrm{p} \right) \notag\\
    Q_\mrm{rad} &= A_\mrm{p} \sigma \epsilon_\mrm{p}\left(T_\mrm{g}^4 - T_\mrm{p}^4\right) \label{eq:fluxes}\\
    Q_\mrm{evap} &= \dot{m}_\mrm{Fe,evap} h_\mrm{Fe,g} + \dot{m}_\mrm{FeO,evap} h_\mrm{FeO,g} \notag
\end{align}

\noindent with {$h_\mrm{p} = \mrm{Nu}\ k_\mrm{f} / d_\mrm{p}$}, the particle heat transfer coefficient, which is determined using the thermal conductivity of the gas-phase with film layer properties $k_\mrm{f}$ and the Ranz-Marshall correlation of the Nusselt number $\mrm{Nu}$, along with the Reynolds $\mrm{Re}$ and Prandtl $\mrm{Pr}$ numbers, similar to Equation \ref{eq:betad}. Radiative heat transfer is only modeled through a simplified Stefan-Boltzmann particle-to-local-gas approximation, with $\sigma=5.67\times10^{-8}\, \SI{}{\watt\meter^{-2}\kelvin^{-4}}$ the Stefan-Boltzmann constant and $\epsilon_\mrm{p}=0.7$ the particle emissivity \cite{wenjian2026}. {The complete effects of radiation in such clustered distributions require the implementation of complex radiative heat transfer models that might be the subject of future work.} The evaporative mass fluxes $\dot{m}_\mrm{Fe,evap}$ and $\dot{m}_\mrm{FeO,evap}$ are modeled similarly to the evaporation model developed by Ramaekers \emph{et al.} \cite{ramaekers2025} as:
\begin{equation}
    \dot{m}_\mrm{Fe,evap} = A_\mrm{Fe} \beta_\mrm{D,Fe} \left( \frac{p_\mrm{vap,1}}{(R_\mrm{u}/W_\mrm{Fe}) T_\mrm{p}}\right)
\end{equation}
\begin{equation}
    \begin{split}
         \dot{m}_\mrm{FeO,evap} &= \frac{A_\mrm{p}}{(R_\mrm{u}/W_\mrm{FeO}) T_\mrm{p}} \times \\ &\quad(\beta_\mrm{D,Fe}\ p_\mrm{vap,2} + \beta_\mrm{D,FeO}\ p_\mrm{vap,3})
    \end{split}
\end{equation}
where $\beta_\mrm{D,Fe}$ and $\beta_\mrm{D,FeO}$ are diffusive mass transfer coefficients of gaseous $\mrm{Fe}$ and $\mrm{FeO}$, calculated using the Sherwood number $\mrm{Sh}$ and the diffusion coefficients of the respective gases, similar to Equation \ref{eq:betad}. $R_\mrm{u}$ is the universal gas constant, and $W_\mrm{\alpha}$ is the molecular weight of species $\alpha$. $p_\mrm{vap,1}$, $p_\mrm{vap,2}$, and $p_\mrm{vap,3}$ are vapor pressures of gaseous $\mrm{Fe}$ in liquid $\mrm{Fe}$, gaseous $\mrm{Fe}$ in liquid $\mrm{FeO}$, and gaseous $\mrm{FeO}$ in liquid $\mrm{FeO}$, respectively. A logarithmic power law approximation for vapor pressures is used:
\begin{equation}
p_\mrm{vap,i}\left(T_\mrm{p}\right) = \exp \left(k_\mrm{i,1} + k_\mrm{i,2} T_\mrm{p}^{-1} + k_\mrm{i,3} \log T_\mrm{p} \right)
\end{equation}
with the model coefficients $k_\mrm{i}$ as in Table \ref{tab:evap}. The evaporated mass of Fe and FeO is subtracted from the particle mass. However, gas-phase Fe and FeO are not tracked explicitly as individual species since the total evaporated mass is negligible.
\begin{table}[h!] \footnotesize
\caption{Logarithmic power law coefficients used in the calculation of vapor pressures of gaseous $\mrm{Fe}$ and $\mrm{FeO}$. Note that the final values are in units atm or $1.01325\times 10^5 \SI{}{\pascal}$.}
\vspace{10pt}
\centerline{\begin{tabular}{cccc}
\hline 
& $k_1$ & $k_2$ & $k_3$\\
\hline
$p_\mrm{vap,1}$   & 35.40 & $-4.963\times10^4$ & -2.433       \\
$p_\mrm{vap,2}$   & 62.08 & $-6.412\times10^4$ & -5.399      \\
$p_\mrm{vap,3}$   & 52.93 & $-6.480\times10^4$ & -4.370       \\
\hline 
\end{tabular}}
\label{tab:evap}
\end{table}
The mass of the particle is updated as follows:
\begin{align}
     \dot{m}_\mrm{Fe} &=  -\nu_\mrm{Fe/O_2}\ \dot{m}_\mrm{O_2} - \dot{m}_\mrm{Fe,evap}\notag \\
    \dot{m}_\mrm{FeO} &= \nu_\mrm{FeO/O_2}\ \dot{m}_\mrm{O_2} - \dot{m}_\mrm{FeO,evap}\label{eq:massp} \\
    \der{m_\mrm{p}}{t} &= \dot{m}_\mrm{Fe} + \dot{m}_\mrm{FeO}\notag
\end{align}
where $\nu_\mrm{Fe/O_2}\approx3.49$ and $\nu_\mrm{FeO/O_2}\approx4.49$ are the stoichiometric mass ratios of the corresponding species for the single-stage reaction, similar to Equation \ref{eq:mr}.

The particle temperature is subsequently solved using a modified Newton-Raphson solver (to account for phase change) over the following equation:
\begin{equation}
    H_\mrm{p} = \frac{m_\mrm{Fe}}{W_\mrm{Fe}} h_\mrm{Fe}(T_\mrm{p}) + \frac{m_\mrm{FeO}}{W_\mrm{FeO}} h_\mrm{FeO}(T_\mrm{p})
\end{equation}

\noindent where $h_\mrm{Fe}$ and $h_\mrm{FeO}$ are the specific enthalpies of $\mrm{Fe}$ and $\mrm{FeO}$ at $T_\mrm{p}$, respectively, which are determined using \textit{NASA} 9-polynomials \cite{mcbride2002nasa}.

The rate of change of particle enthalpy constitutes the source term $S_{H}$ in the gas-phase equations as:
\begin{equation}
    S_\mrm{H} = -\frac{1}{V_{\mrm{cell}}} \sum_{n=1}^{N_p} \der{H_\mrm{p}}{t} \cdot W(\mathbf{x} - \mathbf{x}_p) 
\end{equation}

\subsection{Simulation setup}

In the present work, three groups of simulations were conducted to understand the interplay between the clustering and combustion processes as follows:
\begin{enumerate}
\itemsep0em
    \item Variation in Stokes number $\mrm{St}$
    \item Variation in turbulent Reynolds number $\mrm{Re}_\mrm{\lambda}$
    \item Variation in global equivalence ratio $\phi$ considering FeO as the final oxidation state
\end{enumerate}

The gas phase domain is modeled as a periodic cubical box with a uniform Cartesian grid. The number of computational cells is derived as $N_x = N_y = N_z=\mrm{ceil}(L/\eta)$, where $L$ is the domain size and $\eta=\SI{400}{\micro\meter}$ is the Kolmogorov length. In all simulations, the initial particle and gas temperatures were set at $T_\mrm{p,0} = T_\mrm{g,0} = \SI{1200}{\kelvin}$, the pressure at $p=\SI{1.01325e5}{\Pa}$, with $\YO=0.23$ and $\mathrm{N_2}$ being the only other species in the gas phase. These conditions result in a near-uniform ``ignition" of all particles regardless of clustering, enabling better analysis of the combustion process. Here, the event of ignition is represented by the transition from solid-phase kinetics to the $\mrm{O_2}$-diffusion-limited regime. The particles were initialized as inert monodisperse particles in a Poisson (random homogeneous) distribution in space and allowed to evolve into clusters under the influence of forced turbulence. The stabilized clustered distribution was then allowed to react after $t=\SI{100}{\milli\second}$, which is approximately $2-10\tau_\mrm{c}$, where $\tau_\mrm{c}$ is the cluster formation timescale \cite{hemamalini2025}. The particle and gas-phase data are saved at intervals of approximately $\SI{0.05}{\milli\second}$, and analysis is performed over the entire particle population without sampling. To assess the impact of preferential concentration with a Poisson distribution, the case with $\St=1$, $\Rel=20$, and $\phi=0.75$ was run with the particles starting to react immediately after initialization. Table \ref{tab:stokes2} lists the quantitative parameters corresponding to each case in the three groups of simulations.

The Stokes number $\St$ is evaluated as:
\begin{equation}
    \St = \frac{\tau_\mrm{p}}{\tau_\eta},\quad \tau_\mrm{p}=\frac{\rho_pd_\mrm{p}^2}{18\mu_\mrm{g}}\; \mathrm{and}\; \tau_\eta = \left( \frac{\eta^2}{\nu_g}\right)
\end{equation}
\noindent with $\eta=\SI{400}{\micro\meter}$ the Kolmogorov length of the forced HIT field, $\rho_\mrm{p}$ the particle density, and $\mu_\mrm{g}$ and $\nu_\mrm{g}$ the dynamic and kinematic gas viscosities, respectively.

The global equivalence ratio $\phi$ is determined as follows:
\begin{equation}
    \phi = \frac{m_\mathrm{p,total}}{\rho_\mrm{g} Y_\mrm{O_2} L^3} \cdot \nu_\mrm{O_2/Fe} \label{eq:phi1}
\end{equation}
\noindent where $m_\mrm{p,total}$ is the total $\mrm{Fe}$ mass in the domain, $L$ is the domain length, $\rho_\mrm{g}$ is the density of the gas at $T_\mrm{g}$, and $\nu_\mrm{O_2/Fe}$ is the ratio of molecular weights of $\mrm{O_2}$ and $\mrm{Fe}$, respectively.

\begin{table*}
\renewcommand{\arraystretch}{1.2}
    \footnotesize
\caption{Relevant setup parameters and flow configuration for all cases, with Stokes number $\mrm{St}$, turbulent Reynolds number $\Rel$, equivalence ratio $\phi$, particle sizes $d_\mrm{p}$, domain size $L$, number of particles $n_\mrm{ptcl}$, Kolmogorov length $\eta$, grid resolution $\Delta x$, maximum forced wave number $k_0$, RMS velocity of the forced HIT field $u_\mrm{rms}$. The first row constitutes the reference case, and subsequent rows are arranged based on variations in $\mrm{St}$, $\Rel$ and $\phi$ respectively.}
\vspace{5pt}
\centerline{\begin{tabular}{|c|c|c|c|c|c|c|c|c|c|c|}
\hline 
 Case & $\mrm{St}$ [-]  & $\mrm{Re}_\mrm{\lambda}$ [-] & $\phi$ [-] & $d_\mrm{p}$ [$\SI{}{\micro\meter}$]  & $L$ [cm] & $n_\mrm{ptcl}$ [$\times 10^3$] & $\eta$ [$\SI{}{\micro\meter}$] & $\Delta x$ [$\SI{}{\micro\meter}$] & $k_0$ [$\SI{}{\meter^{-1}}$] & $u_\mrm{rms}$ [$\SI{}{\meter^2/\second}$]\\
\Xhline{1.1pt}
Ref. & 1 & 20 & 0.75 & 10.325 & 2.56 & 647.17 & 400 & 400 & 245 & 0.422 \\
\Xhline{1.1pt}

\multirow{2}{*}{$\mrm{St}$} & 10 & \multirow{2}{*}{20} & \multirow{2}{*}{0.75} & 32.735 & \multirow{2}{*}{2.56} & 20.48 & \multirow{2}{*}{400} & \multirow{2}{*}{400} & \multirow{2}{*}{245} & \multirow{2}{*}{0.422} \\ \cline{2-2} \cline{5-5} \cline{7-7}
 & 50 & & & 73.197 & &  7.68 & & & & \\ \Xhline{1.1pt}

\multirow{2}{*}{$\Rel$} & \multirow{2}{*}{1} & 5 & \multirow{2}{*}{0.75} & \multirow{2}{*}{10.325} & 0.905 & 542.72 & \multirow{2}{*}{400} & 377 & 694 & 0.211 \\
\cline{3-3} \cline{6-7} \cline{9-11}
 & & 10 & & & 1.522 & 557.57 & & 381 & 413 & 0.298 \\ \Xhline{1.1pt}

\multirow{2}{*}{$\phi$} & \multirow{2}{*}{1} & \multirow{2}{*}{20} & 0.25 & \multirow{2}{*}{10.325} & \multirow{2}{*}{2.56} & 216.06 & \multirow{2}{*}{400} & \multirow{2}{*}{400} & \multirow{2}{*}{245} & \multirow{2}{*}{0.422}\\ \cline{4-4} \cline{7-7}
 & & & 0.5 & & & 431.62 & & & & \\
\Xhline{1.1pt}
Poisson & \multicolumn{10}{c|}{ Same as Case Ref. but with a Poisson spatial distribution at start of reaction} \\
\Xhline{1.1pt}
\end{tabular}}
\label{tab:stokes2}
\end{table*}

\subsection{Analysis setup}

\subsubsection{Baseline comparison with 0D constant-volume suspension model}

\begin{figure}[h]
    \centering
    \includegraphics[width=0.4\linewidth]{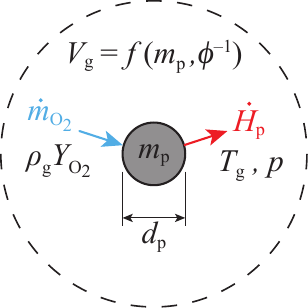}
    \caption{A schematic of the two-way coupled reaction model where the particle of size $d_\mrm{p}$ is coupled to a volume of gas $V_\mrm{g}$ with exchanges in the oxygen from the gas phase into the particle, and heat release from the particle to the gas}
    \label{fig:rxnmodel}
\end{figure}

As periodic boundaries are used in all the simulations presently studied, the domain represents a closed, constant-volume adiabatic box. Hence, as combustion progresses, the gas temperature $T_\mathrm{g}$ increases and, consequently, the pressure $p$ increases. This isochoric process can be simplistically described by a model coupling the 0D reaction model to the gas phase properties. In this regard, a volume of gas $V_\mathrm{g}$ is considered to be coupled to a particle of mass $m_\mrm{p}$ such that the stoichiometric equivalence ratio $\phi$ is given similar to Equation \ref{eq:phi1} as:
\begin{equation}
    \phi = \frac{m_\mathrm{p}}{\rho_\mrm{g} Y_\mrm{O_2} V_\mrm{g}} \cdot \nu_\mrm{O_2/Fe} \label{eq:phi}
\end{equation}
With this construction, $V_\mrm{g}$ can be determined if $\phi$ and the gas and particle properties are known. A schematic of the coupling and the construction is shown in Figure~\ref{fig:rxnmodel}.

The oxygen density in the gas phase $\rho_\mathrm{O_2,g}$ and the gas internal energy $U_\mrm{g}$ are coupled to the particle properties as follows:
\begin{align}
    \der{\rho_\mathrm{O_2,g}}{t} = - \frac{\dot{m}_\mrm{O_2}}{V_\mrm{g}} \label{eq:rho_o2_g}\\
    \der{U_\mrm{g}}{t} = - \der{H_\mrm{p}}{t}
\end{align}
where $\dot{m}_\mrm{O_2}$ denotes the mass consumption of $\mrm{O_2}$ by the particle, and $H_\mrm{p}$ is the particle enthalpy. The gas properties of temperature $T_\mrm{g}$ and pressure $p$ are then evaluated using an iterative Newton-Raphson solver.

\begin{figure}[h]
    \centering
    \includegraphics[width=0.95\linewidth]{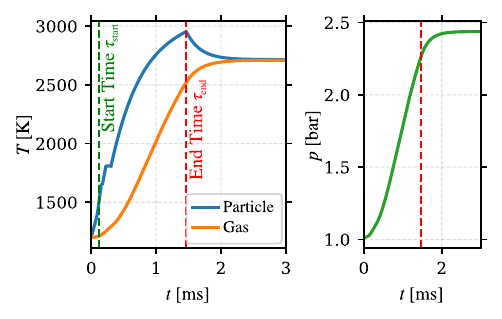}
    \vspace{-10pt}
    \caption{A sample result of the coupled constant volume reaction model with initial particle and gas temperature $T_\mrm{p} = T_\mrm{g} = \SI{1200}{\kelvin}$ for a particle of size $d_\mrm{p} = \SI{10.325}{\micro\meter}$ in air at equivalence ratio $\phi = 0.75$. Left subfigure: Evolution of particle temperature $T_\mrm{p}$ and gas temperature $T_\mrm{g}$ vs. time $t$; Right subfigure: Evolution of gas pressure $p$ vs. time $t$}
    \label{fig:reactionmodel}
\end{figure}

Figure \ref{fig:reactionmodel} shows the evolution of $T_\mrm{p}$, $T_\mrm{g}$ and $p$ over time for a particle of size $d_\mrm{p} = \SI{10.325}{\micro\meter}$ in air with $Y_\mrm{O_2} = 0.2323$, initially at $T_\mrm{p} = T_\mrm{g} = \SI{1200}{\kelvin}$ and $p = \SI{1.01325e5}{\Pa}$, with an equivalence ratio $\phi = 0.75$. This coupled 0D reaction model represents an infinite uniform suspension of particles in space, with each particle coupled only to its "sphere of influence" of gas contained by $V_\mrm{g}$ and no interparticle effects. {\blue In the following sections, this model is referred to as the \textit{0D suspension} model.}

\subsubsection{Characteristic  time scales of iron particle combustion} \label{sec:timescales}

For the analysis of the combustion times in the simulations, the time of transition from solid-phase kinetics to the diffusion controlled regime is considered as the ``start time" $\tau_\mrm{start}$ of combustion, and the time at which all $\mrm{Fe}$ have been consumed is considered as the ``end time" $\tau_\mrm{end}$ of combustion. The time interval between $\tau_\mrm{start}$ and $\tau_\mrm{end}$ is considered as the ``burn time" $\tau_\mrm{B}$. This is annotated in the left subfigure of Figure \ref{fig:reactionmodel}.

\begin{figure}[h]
    \centering
    \includegraphics[width=0.95\linewidth]{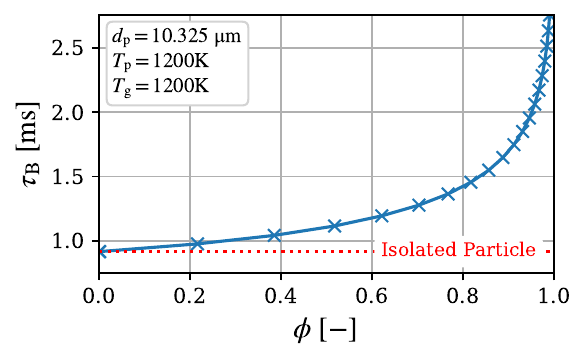}
    \vspace{-10pt}
    \caption{The burn time $\taub$ as a function of the equivalence ratio $\phi$ calculated using the coupled reaction model compared to $\taub$ from the single particle model for the same particle and gas properties as in Figure \ref{fig:rxnmodel}.}
    \label{fig:rxnmodel_taub}
\end{figure}

An analysis of the model at various $\phi$ shows hyperbolic growth of $\taub$ with $\phi$ as shown in Figure \ref{fig:rxnmodel_taub}. At higher values of $\phi>0.8$, $\taub$ is several times higher than the conventional $\taub$ of a single particle reaction model. However, in the range $0<\phi<0.8$, the extension of $\taub$ is approximately linear, inferring that $\taub$ is within the same order of magnitude as the value for a single isolated particle. The burn time statistics for all the simulations are normalized with the corresponding value from the 0D suspension model as $\taub^*$.

Recently, Hemamalini \emph{et al.} \cite{hemamalini2025} derived estimations for the combustion timescale $\taub$ and the cluster evolution timescale $\tau_\mrm{E}$ using a simple first-principles approach. For the case with $\St=1$, $\Rel=20$ and $\phi=0.75$, the particle relaxation timescale is estimated at $\tau_\mrm{p} \approx \SI{1}{\milli\second}$, the combustion timescale at $\tau_\mrm{B} \approx \SI{1.44}{\milli\second}$ and the cluster evolution timescale at $\tau_\mrm{E} \approx \SI{10.99}{\milli\second}$.The order of magnitude difference between $\taub$ and $\tau_\mrm{E}$ indicates that for the conditions simulated in this work, a near-frozen particle distribution during the combustion process is expected, where particle migration has negligible effects and the effects of preferential concentration are maximized.

\subsubsection{Characterization of particle clustering}

There are numerous ways to (qualitatively and quantitatively) analyze particle clustering. Monchaux \emph{et al.} \cite{Monchaux2012} review all the techniques commonly used in the evaluation of particle clustering. In the present work, Vorono\"i decomposition is used to quantify the clustering.

\begin{figure}[h]

    \centering
    \includegraphics[width=\linewidth]{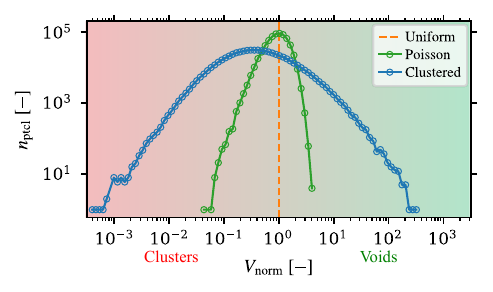}
        
    \vspace{-10pt}
    \caption{Histogram of normalized Vorono\"i volumes $V_\mrm{norm}$ for a uniform suspension, Poisson distribution and fully clustered distribution of particles at $\Rel=20$.}
    \label{fig:voronoicomparison}
\end{figure}

Th\"ater \emph{et al.} \cite{Thaeter2024, thater2026interaction} used Vorono\"i decomposition to quantify particle clustering. To estimate local statistics of particle clustering, Vorono\"i decomposition is used to calculate the Vorono\"i volume $V$--the region of space that is closer to the associated particle than to other particles--associated with each particle. Vorono\"i decomposition is highly beneficial in quantifying local statistics as it incorporates not just one nearest neighbor but an ensemble of particles. In this work, we employ the normalized Vorono\"i volume as
\begin{equation}
  V_\mathrm{norm} = \frac{V}{\bar{V}}
\end{equation}
where $\bar{V}$ is the mean Vorono\"i volume, which is equated to $\bar{V} = L^3/n_\mrm{ptcl}$. Additionally, to quantify the magnitude of clustering, a clustering index similar to that of Monchaux \emph{et al.} \cite{Monchaux2012} is used, utilizing the normalized standard deviation of $V$ as $\sigma(V)/\bar{V}$. For a Poisson distribution, the analytical value of the standard deviation of the normalized volumes is $\sigma(V)/\bar{V}=0.4231$ (see Lazar \emph{et al.} \cite{lazar2013statistical}), with a higher value indicating a clustered distribution. Figure \ref{fig:voronoicomparison} shows the comparison of histogram of normalized Vorno\"i volumes for a Poisson (random homogeneous) distribution, a fully clustered distribution corresponding to $\Rel=20$, and a uniform suspension equivalent to the 0D suspension model. Smaller values of $V_\mrm{norm}$ correspond to dense ``clusters", and conversely, larger values correspond to particle-scarce ``voids".

\subsection{Validation of $\mrm{St}$ and $\Rel$ dependence}

\begin{figure}[h]
    \centering
    \includegraphics[width=0.95\linewidth]{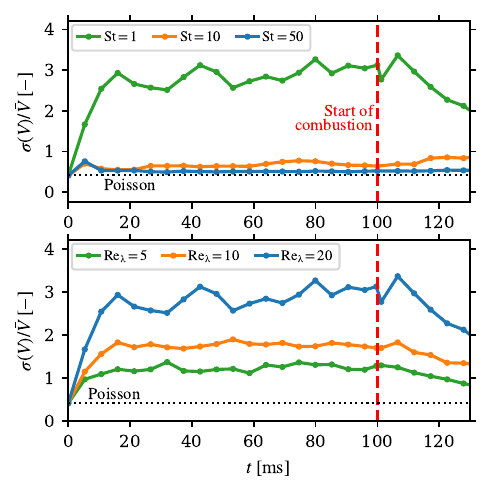}
    \vspace{-10pt}
    \caption{\blue Evolution of the clustering index $\sigma(V)/\bar{V}$ differentiating the trends shown by different $\St$ (top) and $\Rel$ (bottom). Refer to Table \ref{tab:stokes2} for the list of parameters.}
    \label{fig:stokes_meanymin}
\end{figure}

The significance of $\St$ on the magnitude of clustering is well known in the literature; \emph{i.e.}, preferential concentration at the Kolmogorov scale is amplified when the Stokes number, defined with the Kolmogorov timescale, is unity \cite{Wang1993}. The present simulations are consistent with this theory, as in Figure \ref{fig:stokes_meanymin}.

In the present work, only $\St=1$ shows a good magnitude of clustering, \emph{i.e,} a significant increase in $\sigma(V)/\bar{V}$ from the analytical values, as shown in Figure \ref{fig:stokes_meanymin}, owing to the small $\Rel$ considered, thereby validating the phenomenon of preferential concentration at low $\Rel$. Also seen in Figure \ref{fig:stokes_meanymin} is the change in the magnitude of clustering with the combustion process. The value of $\sigma(V)/\bar{V}$ decreases, indicating a change in $\St$ due to the change in particle and gas properties. 

The turbulent Reynolds number $\Rel$ is reported to enhance clustering, as stated by Sumbekova \emph{et al.} \cite{Sumbekova}. A similar trend is observed in the present simulations, as seen in Figure \ref{fig:stokes_meanymin}. It should be noted that $\Rel$ alters the magnitude of the clustering but preserves the timescales of cluster stabilization, as evident from the evolution of $\sigma(V)/\bar{V}$ for different $\Rel$ at the start of the simulation.

\section{Results \& Discussion\label{sec:results}}

\subsection{Visualisation}

\begin{figure}[h]
    \centering
    \includegraphics[width=\linewidth]{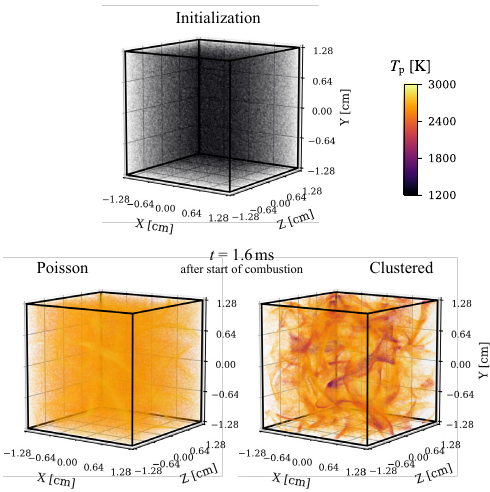}
    \vspace{-10pt}
    \caption{Snapshots of particle distribution colored by particle temperature $T_\mrm{p}$ at initialisation (top) and $\SI{1.6}{\milli\second}$ after the start of combustion (bottom), with a Poisson distribution (bottom left) and a fully clustered distribution at $\St=1$, $\Rel = 20$, and $\phi=0.75$ with $d_\mrm{p} = \SI{10.325}{\micro\meter}$.}
    \label{fig:vis}
\end{figure}

Figure \ref{fig:vis} shows a visualization of the particle distribution field and particle temperatures at initialization and during combustion for a Poisson distribution and a fully clustered distribution. The structure of the preferentially concentrated particle clusters and the inhomogeneity in particle temperature $T_\mrm{p}$ can be clearly observed, whereas the combustion of the Poisson distribution is homogeneous. 3D visualizations of particle temperature $T_\mrm{p}$ and Fe mass fraction $m_\mrm{Fe}/m_\mrm{p}$ are added as supplementary material in the form of videos: \texttt{temperature.mp4} and \texttt{femass.mp4}.

\subsection{Comparison of temperature evolution and combustion times} \label{sec:reynolds}

To assess the progress of combustion across the different scenarios, the comparison of the mean particle and gas temperatures, $T_\mrm{p}$ and $T_\mrm{g}$, respectively, and the burnout times is evaluated.

\begin{figure}
    \centering
    \includegraphics[width=\linewidth]{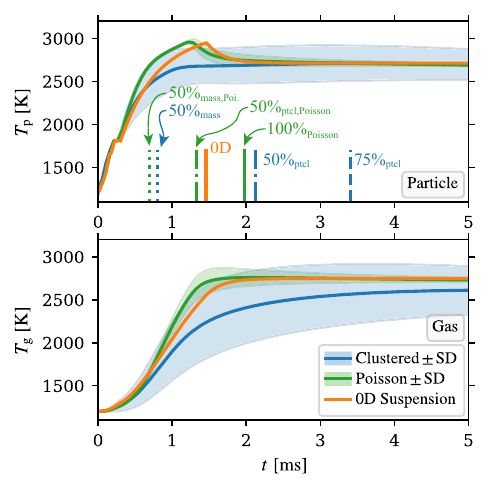}
    \vspace{-10pt}
    \caption{Mean particle (top) and gas (bottom) temperatures in simulations of a clustered (blue in the figure, see Case Ref. in Table \ref{tab:stokes2}) and Poisson (orange in the figure, see Case Poisson in Table \ref{tab:stokes2}) distribution compared to the 0D suspension model (green). Marked in the plot are the combustion end time for the 0D suspension model. For the simulations, the time taken to burnout 50\% of initial $\mrm{Fe}$ mass $50\%_\mathrm{mass}$ and to burn out 50\% and 75\% of particles $50\%_\mathrm{ptcl}$ and $75\%_\mathrm{ptcl}$ are annotated. In the case of Poisson distribution, the total burnout time $100\%_\mathrm{Poisson}$ is also annotated.}
    \label{fig:temp_poisson}
\end{figure}

Figure \ref{fig:temp_poisson} compares the evolution of the mean $T_\mrm{p}$ and $T_\mrm{g}$ in a clustered and a Poisson simulation to the 0D suspension model. The simulation with a Poisson distribution exhibits the maximum mean particle temperature $T_\mrm{p}$ and gas temperature $T_\mrm{g}$. The simulation with the clustered distribution exhibits a slower and smoother transition to equilibrium conditions, owing to the slower combustion mode in clusters, similar to Tha\"ter \emph{et al.} \cite{thater2026interaction}.

From the burnout times annotated in Figure \ref{fig:temp_poisson}, it can be observed that in both the Poisson distribution and the clustered distribution, 50\% of the particles exhibit comparable combustion times to those of the 0D suspension. It is important to note Figure \ref{fig:voronoicomparison} at this point since both the Poisson and clustered distributions exhibit variance in the Vorono\"i volumes that indicate a level of inhomogeneity. Particles that possess $V_\mrm{norm}>1$ might burn faster than predicted by the 0D suspension model, as they effectively burn at a lower $\phi$. For the Poisson distribution, this subtle inhomogeneity is hypothesized to be the cause of the earlier and higher peak temperatures $T_\mrm{p}$ and $T_\mrm{g}$. The preferentially concentrated distribution, on the other hand, has the highest inhomogeneity. 

Hence, the research question \ref{item:rq2} can be answered as follows. The total combustion time of a clustered distribution is significantly longer than that of a Poisson distribution. However, the burnout time of 50\% of the particle population in the clustered distribution is comparable to the combustion times of the Poisson distribution as well as the 0D suspension model, and only a fraction of the particle population has significantly extended combustion times.

\subsection{Effects of $\phi$ on the temperature evolution and combustion times} \label{sec:equivalence}

\begin{figure}[h]
    \centering
    \includegraphics[width=0.95\linewidth]{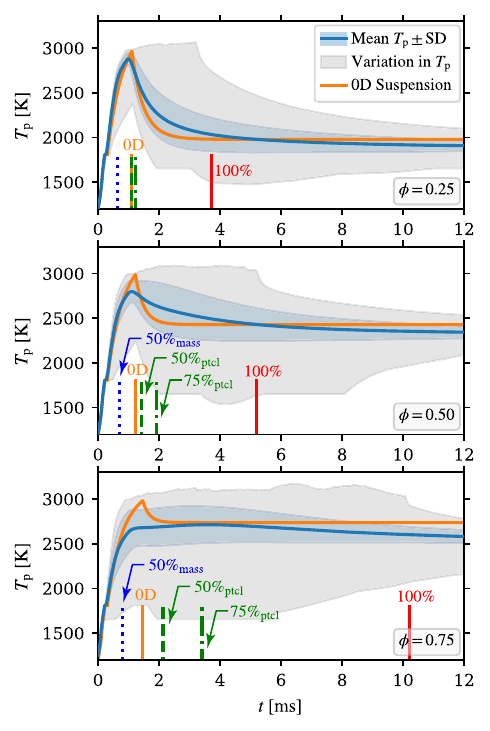}
    \vspace{-10pt}
    \caption{Evolution of mean particle temperature $T_\mrm{p}$ and the observed range compared to the 0D suspension model at different $\phi$ as per Table \ref{tab:stokes2}. Marked in the plot are the combustion end time for the 0D suspension model $\tau_\mrm{E}$ and the time taken to burnout 50\% of initial $\mrm{Fe}$ mass $50\%_\mathrm{mass}$, the time taken to burnout 50\% and 75\% of particles $50\%_\mathrm{ptcl}$ and $50\%_\mathrm{ptcl}$ respectively, and the total burnout time $100\%$ as observed in the simulations.}
    \label{fig:temp_eq}
\end{figure}

Figure \ref{fig:temp_eq} shows the evolution of the particle temperature--the mean and the range--compared to the 0D suspension model at different $\phi$. Even though the evolution of the mean particle temperature resembles the 0D suspension model at lower $\phi$, as shown in Figure \ref{fig:temp_poisson}, the end of combustion is significantly extended. While $\tau_\mrm{end}$ of the 0D suspension model does not show a significant dependence on $\phi$ in the range considered, the combustion end times from the simulations show a strong dependence. Figure \ref{fig:temp_eq} further shows that a majority of the particles still follow combustion modes similar to the 0D suspension model, with half of all the particles completing combustion at or close to $\tau_\mrm{end}$. At $\phi=0.75$, up to 75\% of the particle population finishes combustion within $2\tau_\mrm{end}$. In all the simulated cases, only a fraction ($<25\%$) of the particles burn considerably longer. These results are in line with what was reported by Th\"ater \textit{et al.} \cite{thater2026interaction} in their analysis of the extension of combustion times at various $\phi$.

Hence, the global equivalence ratio $\phi$ significantly affects the total particle burnout times, with an increase of up to a magnitude observed at $\phi=0.75$ compared to $\phi=0.25$, addressing \ref{item:rq3}.

\subsection{Statistical analysis of extension of combustion time}

As stated in the previous sections, half of the particle population in the combustion of a clustered distribution has significantly extended combustion times. This section discusses the characterization of the extension in combustion times of the particles with respect to the spatial characteristics of the particles, \emph{i.e.}, directly correlating the combustion times $\taub$ with the index of preferential concentration $V_\mrm{norm}$. In the following analysis, the normalized combustion time $\taub^*$ with respect to the 0D suspension model is used.

\subsubsection{Effect of $\Rel$ and $\phi$}

As discussed in Sections \ref{sec:reynolds} and \ref{sec:equivalence}, preferential concentration and the subsequent particle clustering significantly extend the combustion times of the particles. First, the spatial characteristics at the start of combustion are compared with the combustion times of each particle.

\begin{figure*}
    \centering
    \includegraphics[width=0.95\linewidth]{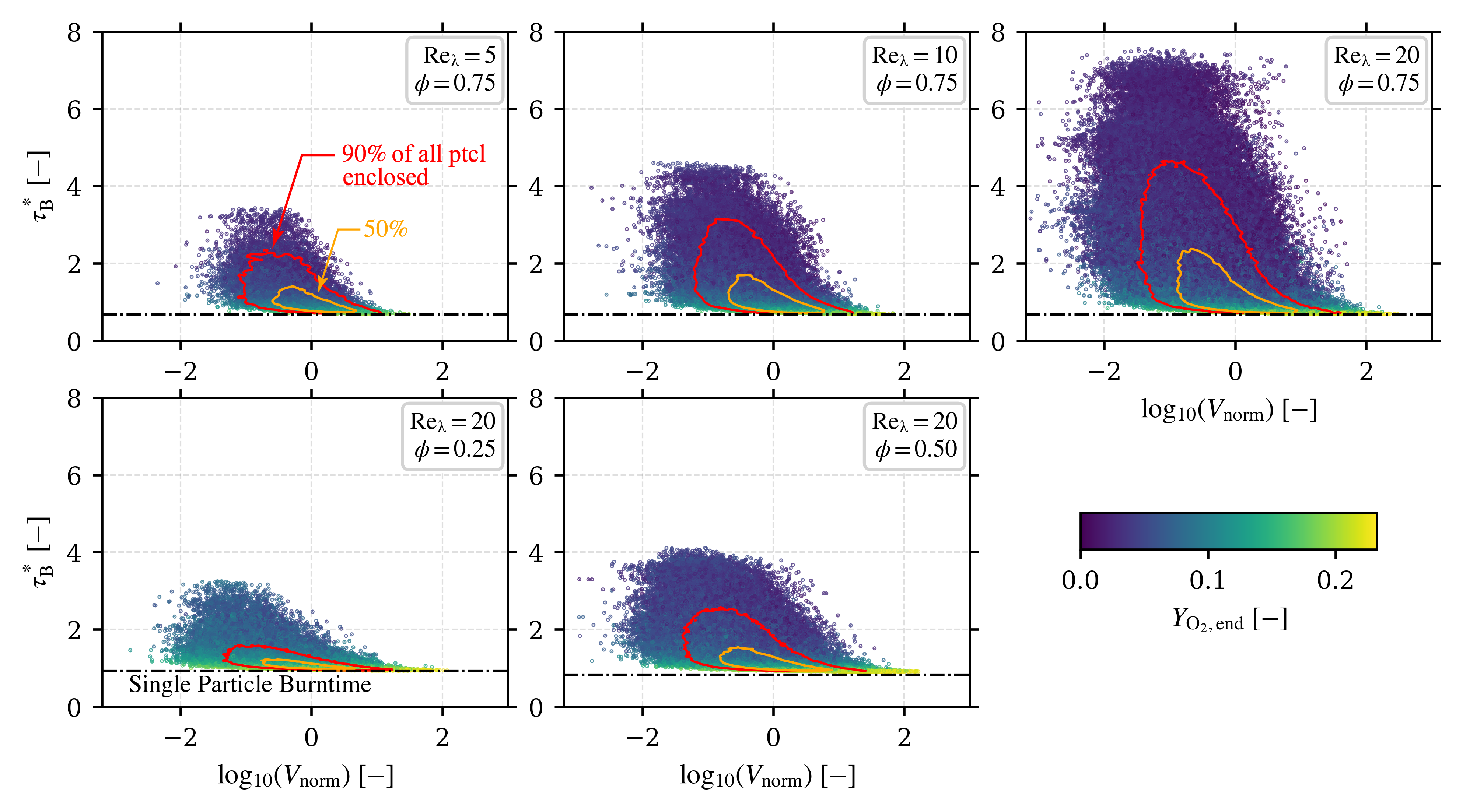}
    \vspace{-10pt}
    \caption{Comparison of the normalized burn time $\taub^*$ in $y-$axis with the normalized Vorono\"i volume of the particles $V_\mrm{norm}$ at the start of combustion $t=t_0$ at different $\Rel$ (top) and $\phi$ (bottom). The points are colored by gas oxygen mass fraction at the location of the particle at the end of combustion $Y_\mrm{O_2,end}$. Refer to Table \ref{tab:stokes2} for the parameters used.}
    \label{fig:burn_re}
\end{figure*}

Figure \ref{fig:burn_re} shows the distribution of normalized combustion time $\taub^*$ (see Section \ref{sec:timescales} for the definition of $\tau_\mrm{B}^*$) with the initial normalized Vorono\"i volume $V_\mrm{norm}$ at the start of combustion, colored by gas $\YO$ at the particle location. Similar to Hemamalini \emph{et al.} \cite{Hemamalini2024}, particles with a longer combustion time $\taub^*$ are well-correlated with more clustered regions. In the present work, an extension of up to eight times was observed for $\Rel = 20$. As shown in Figure \ref{fig:stokes_meanymin}, higher values of $\Rel$ result in stronger clustering magnitudes, which subsequently cause a longer $\taub^*$. 

Particles that have longer combustion times also show lower $\YO$ at the end of combustion. Simulations with a lower $\phi$ for the same $\Rel$ indicate smaller values of $\taub^*$ due to the (relative) surplus of oxygen in the domain. This oxygen surplus at lower $\phi$ also results in a majority of the particles burning close to $\taub^*=1$ for the same $\Rel$, as indicated by the contours in Figure \ref{fig:burn_re}. Hence, the extension of the combustion time is affected by the magnitude of clustering (a consequence of $\Rel$) and also by the available $\mathrm{O_2}$ in the domain (a consequence of $\phi$). 

\subsubsection{Underlying trends in $\taub^*$ vs. $V_\mrm{norm}$}

\begin{figure}[h]
    \centering
    \includegraphics[width=0.95\linewidth]{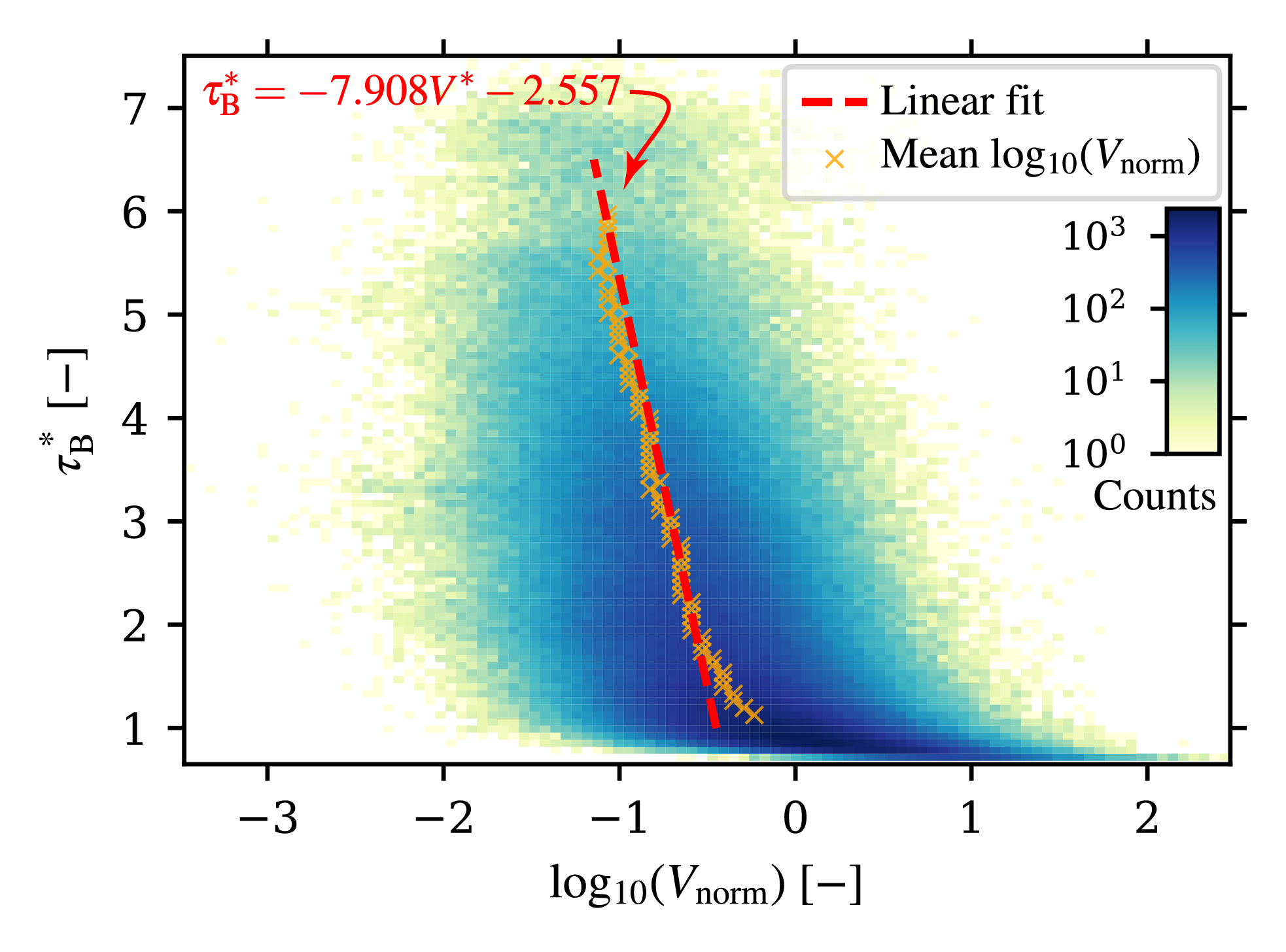}
    \vspace{-10pt}
    \caption{2D histogram of the normalized burn time $\taub^*$ vs. normalized initial Vorono\"i volume $\log_{10}\left(V_\mrm{norm}\right)$ for $\Rel = 20$ and $\phi=0.75$ (see Case Ref. in Table \ref{tab:stokes2}), with the mean of $V_\mrm{norm}$ at each $\taub^*$ for $\log_{10}\left(V_\mrm{norm}\right) < -0.5$ for the cluster region in orange scatter, and a linear fit as a dashed red line. The coefficients of the fit are annotated in the figure.}
    \label{fig:vorpdf}
\end{figure}

\begin{figure}[h]
    \centering
    \includegraphics[width=0.95\linewidth]{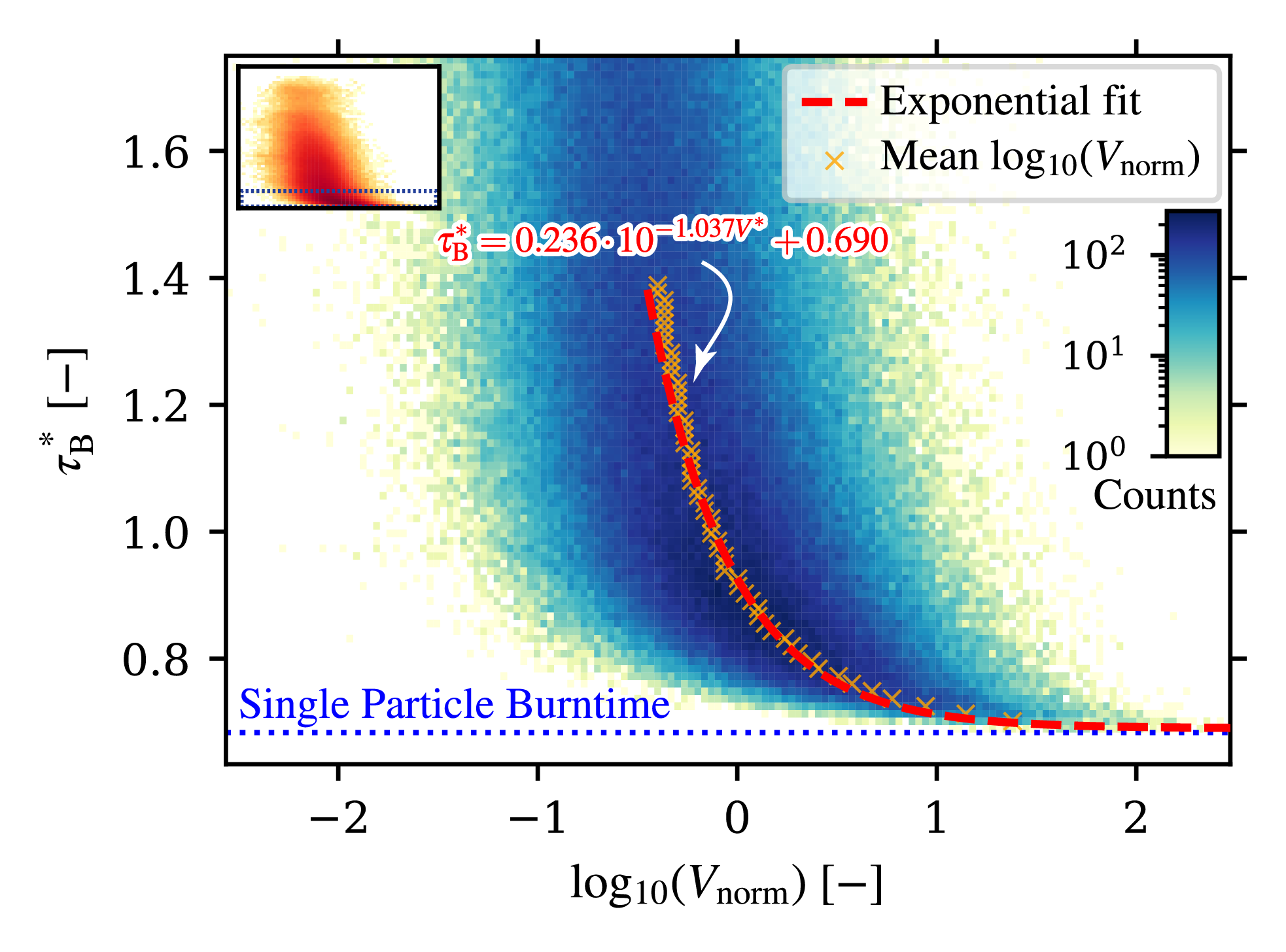}
    \vspace{-10pt}
    \caption{2D histogram of the normalized burn time $\taub^*$ vs. normalized initial Vorono\"i volume $\log_{10}\left(V_\mrm{norm}\right)$ for $\Rel = 20$ and $\phi=0.75$ (see Case Ref. in Table \ref{tab:stokes2}), with the mean of $V_\mrm{norm}$ at each $\taub^*$ for $\log_{10}\left(V_\mrm{norm}\right) > -0.5$ for the void region in orange scatter, and an exponential fit as a dashed red line. The coefficients of the fit are annotated in the figure. Inset shows the full histogram and the clip region to enhance the trend better.}
    \label{fig:burnpdf}
\end{figure}

While Figure \ref{fig:burn_re} shows a scatter plot that might be useful in identifying the range of the parameters considered, a 2D histogram of the distribution is presented in Figures \ref{fig:vorpdf} and \ref{fig:burnpdf}, which shows the density of the scatter distribution. Two remarkable trends--a steep slope in $\log_{10}\left(V_\mrm{norm}\right) < -0.5$ as shown in Figure \ref{fig:vorpdf}, and a flat asymptotic slope in $\log_{10}\left(V_\mrm{norm}\right) > -0.5$ as shown in Figure \ref{fig:burnpdf}-- are also to be noted. These intervals can be attributed as ``clusters'' or particle-dense regions indicated by $\log_{10}\left(V_\mrm{norm}\right) < -0.5$, and ``voids'' or particle-scarce regions indicated by $\log_{10}\left(V_\mrm{norm}\right) > -0.5$. These differing trends are further studied separately.

An analysis of the cluster region, \emph{i.e.}, $\log_{10}\left(V_\mrm{norm}\right) < -0.5$, as shown in Figure \ref{fig:vorpdf}, shows the mean $\log_{10}\left(V_\mrm{norm}\right)$ associated with each $\taub^*$, considering the 1D distribution at each $\taub^*$. Note that the arithmetic mean of the logarithm of a quantity indicates the geometric mean of the quantity, which is logical for a quantity like $V_\mrm{norm}$ that spans multiple orders of magnitude. A direct arithmetic mean may skew the mean towards larger values of $V_\mrm{norm}$. A linear fit is constructed from the collection of points representing the mean $\log_{10}\left(V_\mrm{norm}\right)$ as:
\begin{equation}
    \taub^* \approx -7.908 \cdot \log_{10} (V_\mrm{norm}) - 2.557 \label{eq:vorpdf}
\end{equation}
The slope of $-7.908$ indicates a sharp dependence of $\taub^*$ on $\log_{10}\left(V_\mrm{norm}\right)$ in the particle-dense cluster region of the distribution, implying that particles towards the centers of clusters typically burn logarithmically longer with decreasing $V_\mrm{norm}$, as a direct result of oxygen depletion in the centers of clusters. 

The particle-scarce void region of the distribution, \emph{i.e.}, $\log_{10}\left(V_\mrm{norm}\right) > -0.5$, is isolated, and a similar 2D histogram is presented in Figure \ref{fig:burnpdf}. With this construction, focusing on $\taub^* \approx 1$, the asymptotic trend is easier to identify and fit against the mean $\log_{10}\left(V_\mrm{norm}\right)$ associated with each $\taub^*$ as an exponential function with base 10. The coefficients of the fitted curve are presented as an annotation in Figure \ref{fig:burnpdf}. From the fit, it is estimated that
\begin{equation}
    \taub^* \approx 0.235 \cdot 10^{-1.034 \log_{10}(V_\mrm{norm})} + 0.69 \label{eq:burnpdf}
\end{equation}

Approximating the coefficient in the exponent as -1, a simplified approximation can be made as:
\begin{equation}
    \taub^* \approx 0.69 + \frac{0.235}{V_\mrm{norm}} \label{eq:burnpdf2}
\end{equation}

\noindent indicating a reciprocal function asymptotic to $\taub^* = 0.69$, which is close to the ratio of isobaric and the 0D suspension burn times ($\approx 0.6834$) as discussed in Section \ref{sec:methodology}. The asymptotic behavior at $V_\mrm{norm} \to \infty$ towards the isobaric burn time is expected since at higher $V_\mrm{norm}$, the equivalence ratio localized at the particle location may approach zero, resulting in an isobaric particle-gas-decoupled combustion mode.

Hence, an estimation framework for $\taub^*$ based on the initial $V_\mrm{norm}$ is constructed as:
\begin{equation}
\taub^* \approx 
\begin{cases}
  -7.908 \cdot V^* - 2.557       & \text{if } V^* < -0.5 \\
  0.235 \cdot 10^{-1.034\, V^*} + 0.69      & \text{if } V^* \geq -0.5
\end{cases} \label{eq:caseconstruct}
\end{equation}
where $V^*$ is the abbreviation of $\log_{10}(V_\mrm{norm})$.

While the value of the coefficients may vary with $\Rel$ and $\phi$, it is hypothesized that the cluster and the void region of the particle distribution follow linear and exponential trends, addressing \ref{item:rqa}.

\subsubsection{Can a time-averaged estimation of $V_\mrm{norm}$ minimize the variation in $\taub^*$ vs. $V_\mrm{norm}$?}

\begin{figure}[h]
    \centering
    \includegraphics[width=0.95\linewidth]{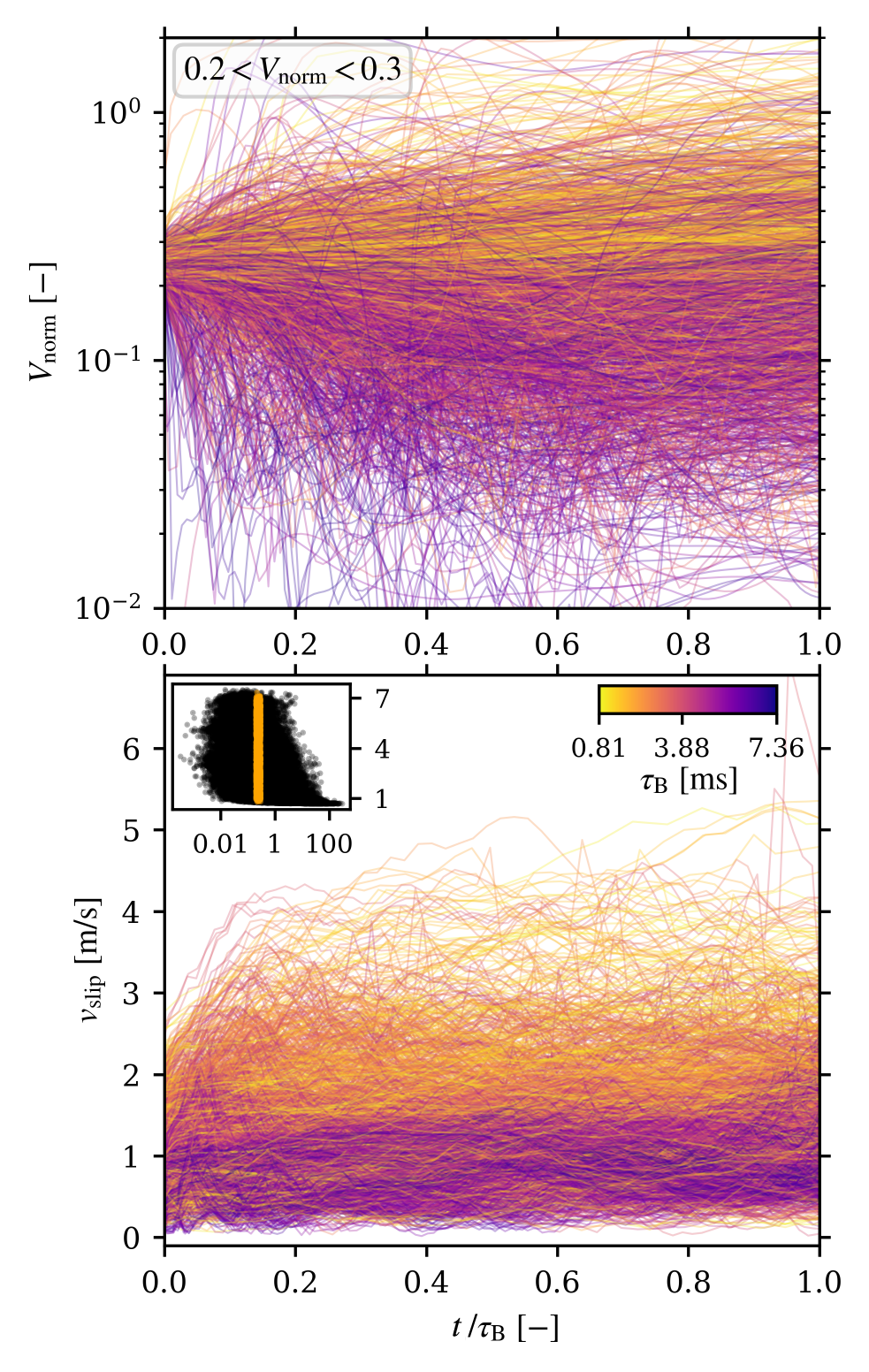}
    \vspace{-10pt}
    \caption{Time history of normalized Vorono\"i volume $V_\mrm{norm}$ (top) and the particle-gas slip velocity $v_\mrm{slip}$ (bottom) for particles that are initially at $V_\mrm{norm} = [0.2,0.3]$ over their combustion time $\taub$  The inset shows the selected range of $V_\mrm{norm}$ from the $V_\mrm{norm}$ vs. $\taub^*$ distribution for $\Rel = 20$ and $\phi=0.75$ as in Figure \ref{fig:burn_re}. The \textit{x}-axis is scaled with $\taub$ of each particle. The lines represent each particle, and the color represents $\taub$ of the particle.}
    \label{fig:timeavg}
\end{figure}

The trends in the cluster and void regions of the particle distribution fit well with the hypothesis that clustering through preferential concentration increases the combustion time of the particles. However, the trends indicated by Figures \ref{fig:vorpdf} and \ref{eq:burnpdf} do not explain the variation in $\taub^*$ for particles at the same $V_\mrm{norm}$ at the start of combustion. Since homogeneous isotropic turbulence and the coupled particle motion are random in the statistical sense, the particles are not necessarily bound to the same $V_\mrm{norm}$ at timescales larger than the particle relaxation timescale $\tau_\mrm{p} \approx \SI{1}{\milli\second}$, as estimated in Section \ref{sec:methodology}. As many particles exhibit a combustion time $\taub$ longer than $\tau_\mrm{p}$, the initial $V_\mrm{norm}$ might not indicate the clustering magnitude over the combustion time of the particles. However, the particle-gas slip velocity (given by the magnitude of the difference in particle and gas velocities $v_\mrm{slip} = |\mathbf{v}_\mrm{p} - \mathbf{v}_\mathrm{g}|$) might indicate the tendency of the particles to deviate from their original location in the local cluster and, hence, is a quantity of interest in monitoring the time history of the particles through combustion.  Figure \ref{fig:timeavg} shows the evolution of $V_\mrm{norm}$ and the particle-gas slip velocity $|v_\mrm{slip}|$ over the combustion times $\taub$ for particles that are initially at $V_\mrm{norm}=[0.2,0.3]$. Note that the abscissa is $t/\taub$ with $t/\taub=0$ and $t/\taub=1$ indicating the start and end of combustion of the particles, respectively. 

\begin{figure}[h]
    \centering
    \includegraphics[width=0.95\linewidth]{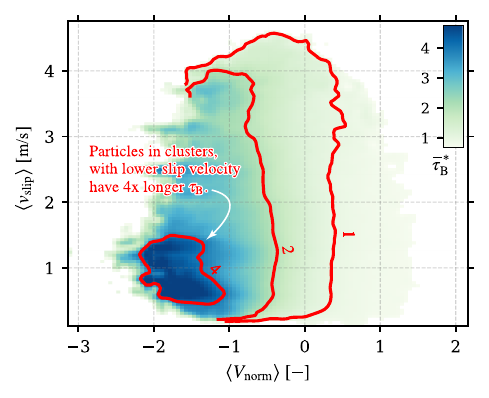}
    \vspace{-10pt}
    \caption{Comparison of the time-average normalized Vorono\"i volume $\langle V_\mrm{norm} \rangle$ and the time-averaged particle-gas slip velocity $\langle v_\mrm{slip} \rangle$ for $\Rel = 20$ and $\phi=0.75$ (see Case Ref. in Table \ref{tab:stokes2}) as a scatter plot colored by the normalized combustion time of each particle $\taub^*$ (top) and a 2D histogram showing the mean $\taub^*$ in each bin (bottom). Contours in the histogram represent $\taub^*=[1,2,4]$ indicating higher $\taub^*$ at lower $\langle V_\mrm{norm} \rangle$ and $\langle v_\mrm{slip} \rangle$. }
    \label{fig:vslip}
\end{figure}

Figure \ref{fig:timeavg} shows a weak correlation of $\taub^*$ with the time history of both $V_\mrm{norm}$ and $v_\mrm{slip}$. Particles tending towards a higher $V_\mrm{norm}$ burn considerably faster, as do particles with a higher $v_\mrm{slip}$. From Figure \ref{fig:timeavg}, it is evident that the initial spatial characteristics may not be fully meaningful in estimating their effects on combustion. For this reason, a simple arithmetic mean of $V_\mrm{norm}$ and $v_\mrm{slip}$ over the combustion time of the particles is computed on a per-particle basis to obtain the time-averaged equivalents of the quantities--$\langle V_\mrm{norm}\rangle$ and $\langle v_\mrm{slip} \rangle$, respectively. The distribution and the corresponding 2D histogram of the time-averaged quantities are presented in Figure \ref{fig:vslip}.

\begin{figure}[h]
    \centering
    \includegraphics[width=0.95\linewidth]{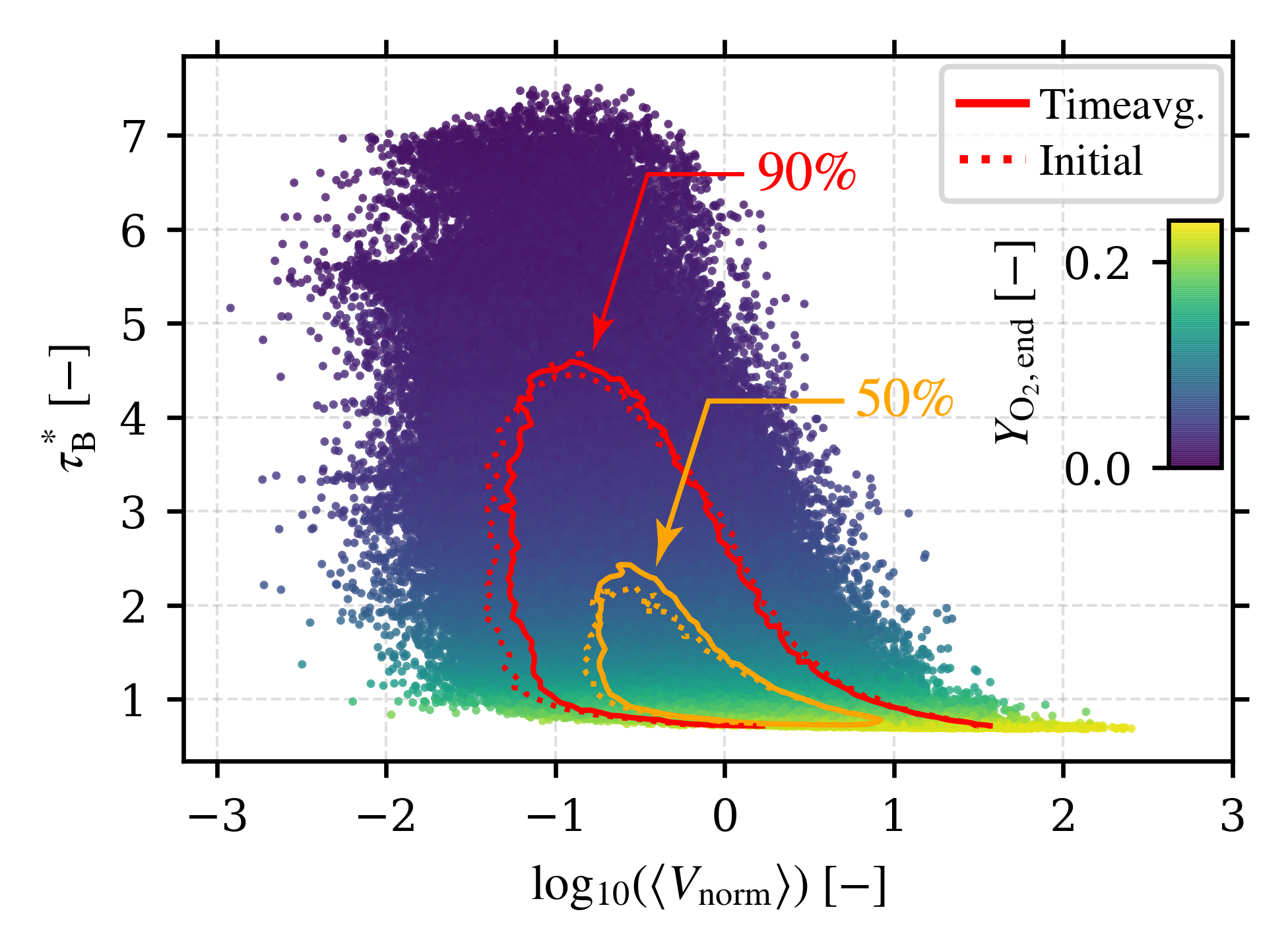} 
    \vspace{-10pt}
    \caption{Comparison of the normalized burn time $\taub^*$ in $y-$axis with the time-averaged normalized Vorono\"i volume of the particles $\langle V_\mrm{norm} \rangle$ in the $x-$axis for $\Rel=20$ and $\phi=0.75$ (see Case Ref. in Table \ref{tab:stokes2}). The points are colored by gas oxygen mass fraction at the location of the particle at the end of combustion $Y_\mrm{O_2,end}$. The contour lines enclose $50\%$ (orange) and $90\%$ (red) of the points. Dotted lines indicate contours of $V_\mrm{norm}$ vs. $\taub^*$ as in Figure \ref{fig:burn_re} and solid lines indicate contours of $\langle V_\mrm{norm} \rangle$ vs. $\taub^*$.  }
    \label{fig:timeavgvnorm}
\end{figure}

Following the trends seen in Figure \ref{fig:timeavg}, which are restricted to a narrow selection of $V_\mrm{norm} = [0.2,0.3]$, Figure \ref{fig:vslip} shows the distribution of $\langle V_\mrm{norm}\rangle$ and $\langle v_\mrm{slip} \rangle$ for all the particles and further adds to the hypothesis that the particles with elongated $\taub$ not only possess a lower value of $\langle V_\mrm{norm}\rangle$ but also, importantly, have lower values of $\langle v_\mrm{slip} \rangle$. 
Although Figure \ref{fig:vslip} gives a deeper insight into the elongation of $\taub^*$, the comparison of $\langle V_\mrm{norm}\rangle$ with the $\taub^*$ shown in Figure \ref{fig:timeavgvnorm} does not show a major improvement over the comparison of the initial value of $V_\mrm{norm}$, as shown in Figure \ref{fig:burn_re}. In fact, the time-averaged value of $V_\mrm{norm}$ only weakly improves the scatter in the cluster region of the data ($V_\mrm{norm} < 1$) and shows no change to the void region of the data ($V_\mrm{norm} > 1$). This preference with $V_\mrm{norm}$ can be attributed to the fact that in the void region of the data, the particles have a (statistically) shorter $\taub$, such that $\taub \sim \tau_\mrm{p}$ and hence, the time-averaged $\langle V_\mrm{norm}\rangle$ over $\taub$ does not result in a considerably different value than the initial $V_\mrm{norm}$. In other words, in the void region, $\langle V_\mrm{norm}\rangle \approx V_\mrm{norm}$. However, in the cluster region of the data, $\taub \ge \tau_\mrm{p}$ such that time-averaging over a longer $\taub$ captures more of the particle motion and leads to a noticeable variation of $\langle V_\mrm{norm}\rangle$ over $V_\mrm{norm}$.
\begin{figure}[h]
    \centering
    \includegraphics[width=0.95\linewidth]{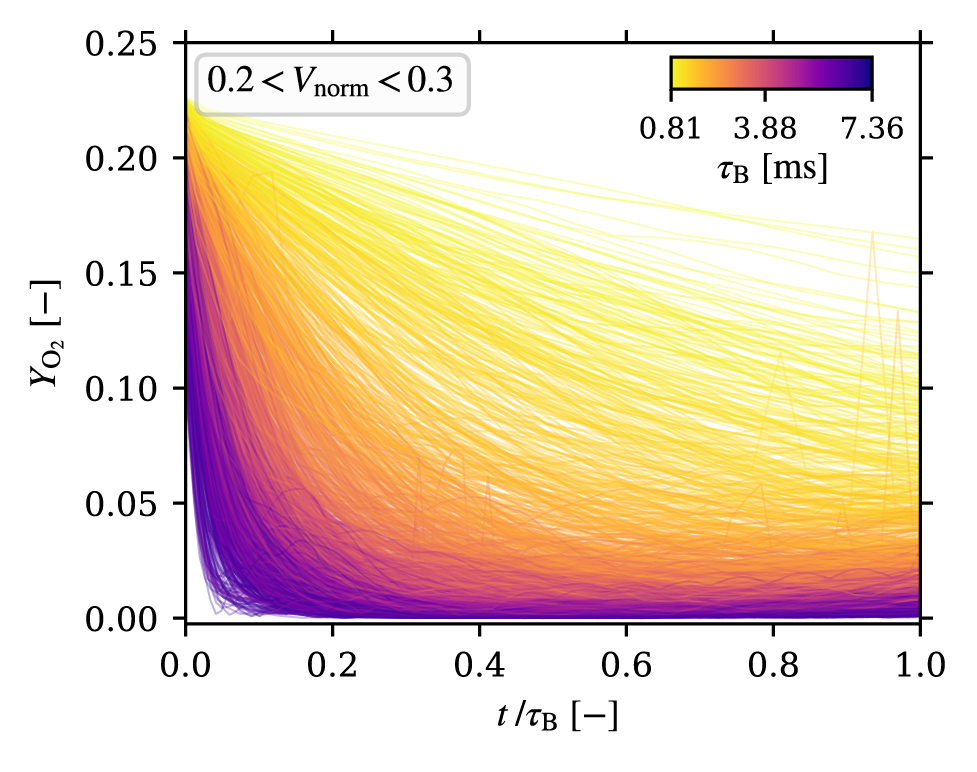}
    \vspace{-10pt}
    \caption{Time history of $\YO$ at the particle location for particles that are initially at $V_\mrm{norm} = [0.2,0.3]$ over their combustion time $\taub$ for $\Rel = 20$ and $\phi=0.75$ (see Case Ref. in Table \ref{tab:stokes2}). The \textit{x}-axis is scaled with the combustion time $\taub$ of each particle. The lines represent each particle, and the color represents the combustion time $\taub$ of the particle.}
    \label{fig:yo2pdf}
\end{figure}

More importantly, time-averaged spatial characteristics on a per-particle basis still exhibit a variation in the estimation of $\taub^*$, answering question \ref{item:rqb}. Whereas comparing the time-history of $\YO$ at the particle location over the combustion time of each particle shows a very strong correlation with the combustion time, as shown in Figure \ref{fig:yo2pdf}. While it is easy to infer that the availability of $\mrm{O_2}$ directly results in the extension of the combustion time $\taub$, the $\mrm{O_2}$ availability (or lack thereof) at the particle location is merely a consequence of spatial characteristics. Figures \ref{fig:timeavg}, \ref{fig:vslip}, \ref{fig:timeavgvnorm}, and \ref{fig:yo2pdf} indicate that the cluster microstructure given by $V_\mrm{norm}$ does not completely explain $\mrm{O_2}$ depletion.

\subsubsection{What other spatial structures can influence $\taub$ extension?}

\begin{figure*}[h]
    \centering
    \includegraphics[width=0.935\linewidth]{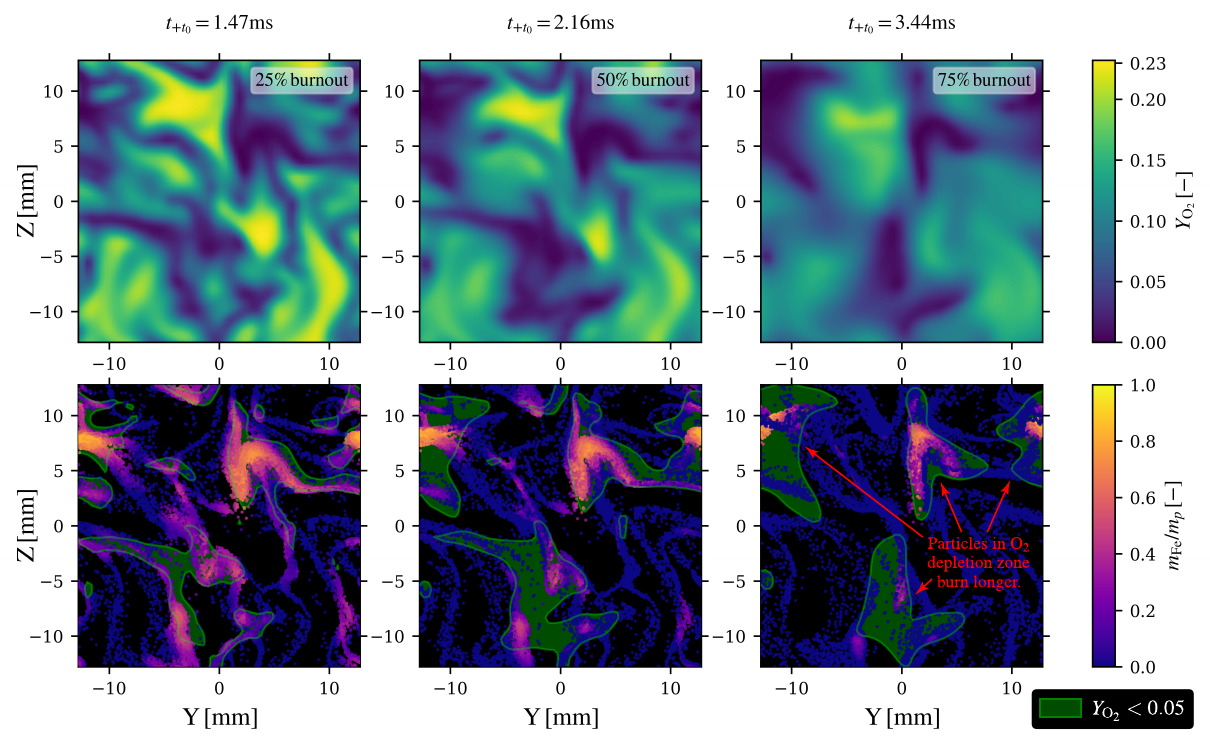}
    \vspace{-10pt}
    \caption{Contours of gas $\YO$ (top) and particle coordinates (bottom) for $\Rel = 20$ and $\phi=0.75$ at a slice of the domain in the $x-$direction at $x=[-8.6,-8.2]\SI{}{\milli\meter}$ (arbitrarily chosen) at simulation time of 1.47, 2.16 and 3.44$\SI{}{\milli\second}$ after start of combustion. The particles are colored by the Fe mass fraction $m_\mrm{Fe}/m_\mrm{p}$. The chosen times correspond to 25\%, 50\% and 75\% particle burnout times. The green contour line in bottom figure correspond to a contour line of gas $\YO=0.05$ indicating (arbitrarily) the $\mrm{O_2}$-depletion zone.}
    \label{fig:slice}
\end{figure*}

While $V_\mrm{norm}$ is an effective parameter to quantify the intra-cluster structure through the spatial particle distribution, inter-cluster structures cannot be fully represented by $V_\mrm{norm}$. Figure \ref{fig:slice} shows the evolution of the $\mrm{O_2}$ depletion zone through the contour of $\YO$, and the corresponding particle distribution with the particle $\mrm{Fe}$ mass fraction ($m_\mrm{Fe}/m_\mrm{p}$). As shown in Figure \ref{fig:slice}, the $\mrm{O_2}$ depletion zone not only depends on the intra-cluster structure (as seen at 25\% burnout time) but also on the proximity of adjacent clusters. The adjacency of multiple clusters cannot be estimated from the intra-cluster structure characterized by $V_\mrm{norm}$. If multiple clusters are adjacent to each other, the combustion time $\taub$ might be extended for the same $V_\mrm{norm}$. Considering that the diffusion of $\mrm{O_2}$ is controlled by both convection and diffusion, the former of which is defined by the HIT field, it is difficult to \textit{predict} the size of the $\mrm{O_2}$ depletion zone and, subsequently, the particles that exhibit an extended combustion time $\taub$ deterministically. Also shown in Figure \ref{fig:slice} with the particle distribution is a contour of $Y_\mrm{O_2} = 0.05$ that indicates that the depletion zone might not necessarily overlap the particle clusters beyond a certain time, and the clusters that do overlap exhibit an extended combustion time.
\begin{figure}[h]
    \centering
    \includegraphics[width=0.925\linewidth]{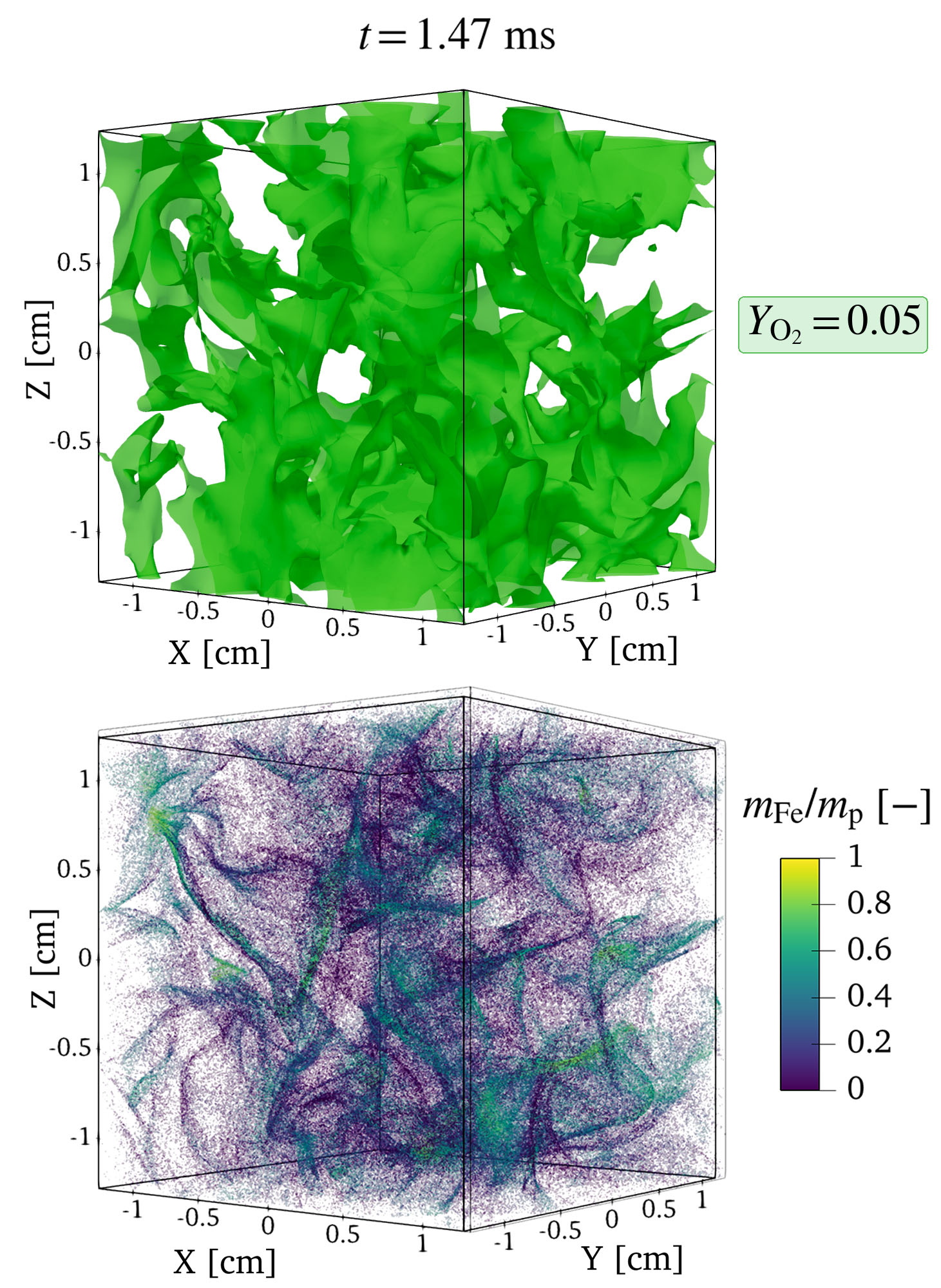}
    \caption{3D visualization of the gas-phase $\YO=0.05$ contour (top) and particle distribution colored by the $\mrm{Fe}$ mass fraction $m_\mrm{Fe}/m_\mrm{p}$ (bottom) at 25\% burnout time at $\Rel=20$ and $\phi=0.75$ (see Case Ref. in Table \ref{tab:stokes2}).}
    \label{fig:visyo2femass}
\end{figure}

Figure \ref{fig:visyo2femass} shows the 3D structure of the $\mathrm{O_2}$ depletion zone, simplified as the contour $\YO = 0.05$, and the particle distribution colored by the $\mrm{Fe}$ mass fraction. As the clusters have a 3-dimensional dynamic structure resembling strings of particles rather than a spherical structure, methods to quantify the cluster size and subsequently the size of the $\mrm{O_2}$ depletion zone, such as radial distribution functions (RDF) or box-counting may be arbitrary. This string-like structure may also aid the (molecular) diffusion of $\mrm{O_2}$ inwards, as molecular diffusion is isotropic. This has also been elucidated by Tha\"ater \emph{et al.} \cite{thater2026interaction}, who observe a narrowing of the clusters as a result of this diffusion-driven flow. Hence, the \textit{prediction} of combustion time $\taub$ from just the initial particle distribution and the initial particle and flow characteristics is only possible to the extent of estimation, as shown in Figures \ref{fig:vorpdf} and \ref{fig:burnpdf}, and a variation in $\taub$ is always to be expected; \textit{i.e.,} \textit{accurate prediction of $\taub$ is not possible in a deterministic sense} solely based on the intra-cluster structure at the onset of the combustion process. Rather, $\taub$ of each particle is a result of its complex individual and collective position and motion via $\mrm{O_2}$ depletion zones in the gas phase over the course of combustion, answering \ref{item:rqc}.

\begin{figure}[h]
    \centering
    \includegraphics[width=0.95\linewidth]{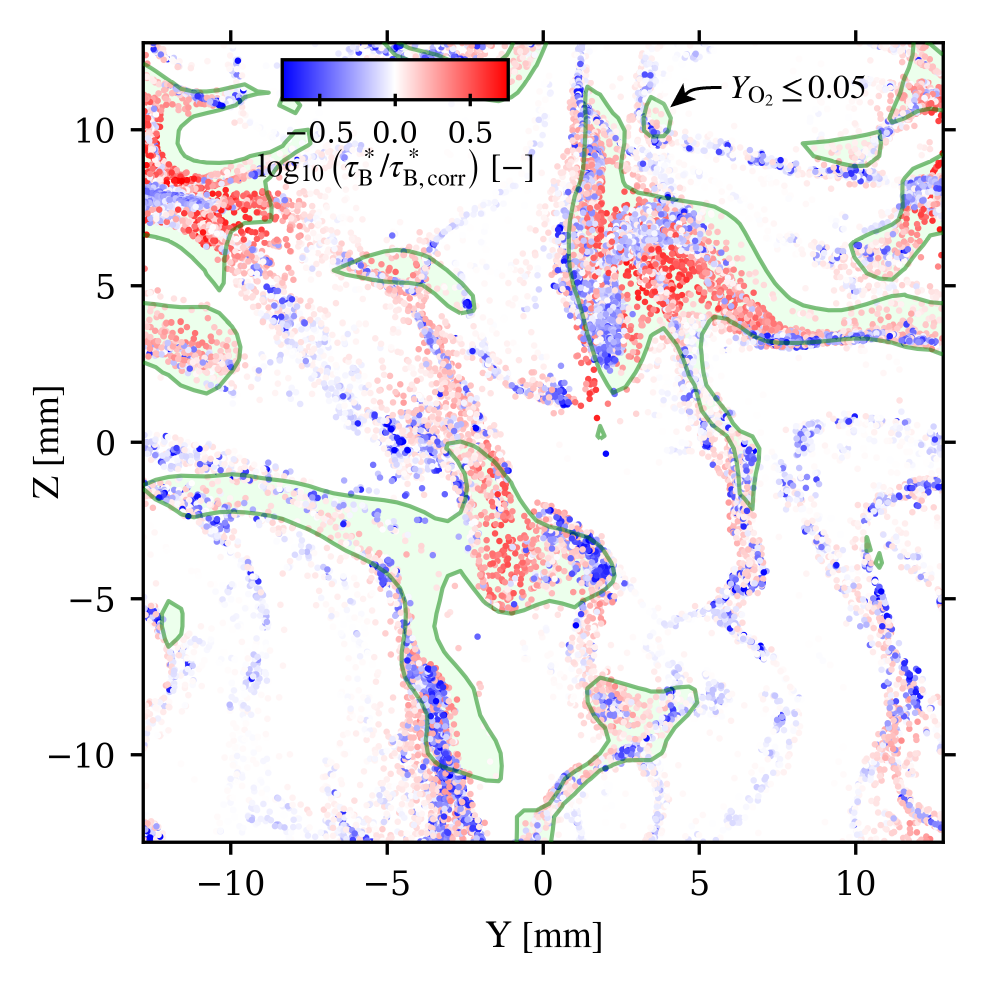}
    \vspace{-10pt}
    \caption{The particles as in Figure \ref{fig:slice} colored by the ratio between the normalized combustion time from the simulation $\taub^*$ and the prediction $\tau^*_\mrm{B,corr}$ from Equation \ref{eq:caseconstruct}. $\log_{10}(\taub^*/\tau^*_\mrm{B,corr}) > 0$ indicates under-prediction by the model meaning the particles burn longer than predicted, and vice versa. This snapshot corresponds to the case of $\Rel=20$ and $\phi=0.75$ (see Case Ref. in Table \ref{tab:stokes2}).}
    \label{fig:correlation}
\end{figure}

Figure \ref{fig:correlation} shows the comparison of the combustion time of the particles shown in Figure \ref{fig:slice} with the estimation from the framework as in Equation \ref{eq:caseconstruct}. Please refer to the supplementary material: \texttt{correlation.mp4} for a three-dimensional animation of the quantities in Figure \ref{fig:correlation} for a global perspective. The index presented in Figure \ref{fig:correlation} represents the mismatch of the combustion time of the particles with the framework--a positive value indicates a higher combustion time than predicted, and vice versa. Higher-than-predicted combustion times correspond to particles in a region with multiple clusters and correlate well with the $\mrm{O_2}$ depletion zone, as shown in Figure \ref{fig:slice}. Particles with overpredicted combustion times occur in clusters that are not adjacent to other clusters. Hence, the framework in Equation \ref{eq:caseconstruct} underpredicts combustion time for cluster-dense regions with multiple adjacent clusters and overpredicts for singular isolated clusters. Since the adjacency of clusters--inter-cluster structure--cannot be estimated by $V_\mrm{norm}$, this prediction bias is as expected. Further analysis of the collective combustion behavior of clustered iron particles is essential for developing quantifiable parameters that characterize inter-cluster structures—such as cluster size and spacing—and their influence on combustion duration.

\section{Summary of findings \label{sec:conclusion}}

In the present work, gas-phase DNS of turbulent iron particle combustion in a forced HIT field is conducted to better understand the complex relationship between turbulence and the combustion process. The following results are presented and discussed in this work:
\begin{enumerate}[label=RQ:\Roman*,leftmargin=1.1cm, wide]
\itemsep0em
    \item A visual comparison of the combustion process in a preferentially concentrated particle distribution with a Poisson distribution shows strong inhomogeneity in the particle and gas properties owing to the formation of densely and sparsely populated regions. The mean temperature evolution of the clustered particle distribution is slower and has a significantly lower peak value due to the depletion of $\mrm{O_2}$ in particle-rich clusters, resulting in a slower combustion mode. 
    \item Increasing $\phi$ also leads to an extension of burn times as a consequence of localized depletion of $\mrm{O_2}$, in agreement with the results observed by Tha\"ter \textit{et al.} The combustion time also increases with increasing $\Rel$ due to the increased magnitude of clustering at higher $\Rel$. Particles with longer burn times $\taub^*$ are significantly correlated with lower values of $V_\mrm{norm}$, as well as with lower values of $\YO$ at the end of combustion. 
    \item A deterministic prediction of the combustion time $\taub^*$ (relative to a 0D suspension model) based on the initial cluster structure given by $V_\mrm{norm}$ is only possible to the extent of the correlation:
    \begin{equation}
    \taub^* \approx 
    \begin{cases}
      -7.908 \cdot V^* - 2.557       & \text{if } V^* < -0.5 \\
      0.235 \cdot 10^{-1.034\, V^*} + 0.69      & \text{if } V^* \geq -0.5
    \end{cases} 
    \end{equation}
    \noindent where $V^*$ is the abbreviation of $\log_{10}(V_\mrm{norm})$. This correlation is approximated for $\Rel=20$ and $\phi=0.75$. However, a similar logarithmic and asymptotic correlation is expected for other values. This finding is supported by the following statements of RQ:IV-VI.
    
    \item Analyzing the trends of $V_\mrm{norm}$ vs. $\taub^*$ with respect to particle-dense and particle-scarce regions of the distribution indicates an exponential increase in combustion time with decreasing $V_\mrm{norm}$ in the cluster region, indicating that particles at the center of clusters have elongated $\taub^*$, and an asymptotic correlation in the void region means that particles in voids burn in an uncoupled combustion mode similar to isobaric conditions.
    \item The time-history of particles through their combustion process offered insight into the significance of $v_\mrm{slip}$--particles with lower $v_\mrm{slip}$ and lower $V_\mrm{norm}$ statistically have a longer $\taub^*$, indicating lesser prevalence to move out of their initial cluster. However, time-averaged $V_\mrm{norm}$ shows minimal improvement in the correlation with $\taub^*$, implying that the microstructure alone is incapable of explaining the extension in combustion time.
    \item Visualization of the evolution of gas $\mrm{O_2}$ depletion zone and the corresponding particle distribution indicates the significance of the macrostructure. Adjacent clusters might yield a larger depletion zone which can extend $\taub^*$ irrespective of the microstructure. This is verified by mapping particles with longer $\taub^*$ than predicted by the trends in microstructure. 
    
\end{enumerate}

\subsection{Limitations and outlook}

The simulations in this study are DNS with forced turbulence, which may not completely match realistic flows. However, the simulations allow for the analysis of the interplay between turbulence and combustion in a controlled setting to provide insight into the research problem from a fundamental physical perspective. To further extend the research--in the direction of physics as well as realistically accurate conditions-- experimental input regarding the properties of turbulence and the thermophysical properties in a turbulent iron combustor is vital. Furthermore, simulations with stronger $\Rel > 100$ may be conducted to analyze the multi-scale clustering independent of $\mrm{St}$. This work also utilizes a simplified radiative heat transfer model. Further effort to isolate and quantify the effects of radiative heat transfer, such as intra-cluster, inter-cluster, and cluster-to-void, is required.

In realistic cases, the particle distribution is expected to be polydisperse, with a range of particle sizes (at a range of $\St$). Preferential concentration in polydisperse flows is expected to be generally weaker \cite{Aliseda}; however, this also depends on the range of sizes considered \cite{Sumbekova}. Another variable in such polydisperse distributions is the variation in combustion timescales, which roughly scales as $d_\mrm{p}^2$ \cite{hemamalini2025}. Extensive simulations regarding the effects of a polydisperse particle distribution on the clustering process, and subsequently, the combustion process of the particles are necessary. 

\section*{CrediT authorship contribution statement}

SSH - Conceptualization, Methodology, Visualization, Formal analysis, Investigation, Writing - original draft \& editing. BC - Conceptualization, Methodology, Software, Writing-review. XCM - Project administration, Conceptualization, Funding acquisition, Investigation, Supervision, Writing-review. 

\section*{Declaration of competing interest}

The authors declare that they have no known competing financial
interests or personal relationships that could have appeared to
influence the work reported in this paper.

\section*{Acknowledgments}

This project has received funding from the Eindhoven Institute of Renewable Energy Systems (EIRES) Start-Up Package (CRT STA MS-FeComb). This work used the Dutch national e-infrastructure with the support of the SURF Cooperative with the grant number 2024.003. This work was sponsored by NWO - Domain Science for the use of supercomputer facilities. 
\FloatBarrier

\bibliographystyle{cnf-num}
\bibliography{cnf-refs.bib}

{\red
\section*{Appendix}

\subsection*{A - Implementation of HIT Forcing}

To maintain a statistically stationary turbulent field within the periodic domain, a stochastic forcing method is employed. The turbulence is sustained by injecting energy into the low-wavenumber modes ($k \leq k_f$), allowing the natural energy cascade to develop toward the smaller dissipative scales. For each targeted mode, a complex acceleration coefficient $\mathbf{a}(\mathbf{k}, t)$ is evolved using an Ornstein-Uhlenbeck process, which ensures temporal correlation of the forcing:
\begin{equation}
    \mathbf{a}(\mathbf{k}, t + \Delta t) = \mathbf{a}(\mathbf{k}, t) \left( 1 - \frac{\Delta t}{\tau_\mrm{f}} \right) + \sigma_\mrm{f} \sqrt{\frac{2 \Delta t}{\tau_\mrm{f}}} \mathbf{\Psi}
\end{equation}
where $\tau_\mrm{f}$ is the correlation time, $\sigma_\mrm{f}$ is the forcing amplitude, and $\mathbf{\Psi}$ is a vector of Gaussian random numbers. The coefficients are projected to be divergence-free, ensuring $\nabla \cdot \mathbf{f} = 0$ in physical space as:
\begin{equation}
    \mathbf{b}(\mathbf{k}, t) = \mathbf{a}(\mathbf{k}, t) - \frac{\mathbf{a}(\mathbf{k}, t) \cdot \mathbf{k}}{|\mathbf{k}|^2} \mathbf{k}
\end{equation}
The resulting forcing acceleration $\mathbf{f}(\mathbf{x}, t)$ is obtained via inverse Fourier transform with $\mathbf{b}(\mathbf{k}, t)$ as the amplitude for wavenumber $\mathbf{k}$. The resulting physical acceleration field $f_i$ is then used to define the source terms in Equations \ref{eq:momentum} and\ref{eq:energy}:
\begin{align}
    S_{\mrm{turb},i} &= \rho f_i \\
    S_{\mrm{turb},H} &= \rho u_i f_i - e_{inj}
\end{align}
The term $e_{inj}$ represents the volume-averaged energy injection rate, which is subtracted to maintain global energy conservation and prevent artificial temperature drift. In the current work, the Kolmogorov length is fixed for all the simulations at $\eta=\SI{400}{\micro\meter}$. Thus, the turbulent dissipation rate $\epsilon$ can be estimated as:
\begin{equation}
    \epsilon = \nu_\mrm{g}^3/\eta^4 \label{eq:turbeps}
\end{equation}
where $\nu_\mrm{g}$ is the kinematic viscosity of the gas. Next, the turbulent Reynolds number $\mathrm{Re}_\mrm{\lambda}$ is assumed a value, which enables the computation of the fundamental wave number $k_0$, and subsequently, the domain length $L=2\pi/k_0$ from the relation:
\begin{equation}
    \mathrm{Re}_\mrm{\lambda} = \epsilon^{1/3}k_0^{-4/3}/\nu_\mrm{g} \label{eq:rel}
\end{equation}
Eswaran and Pope \cite{Eswaran1988} provide the following analytical solution for the dissipation rate $\epsilon_\mathrm{th}$ as a result of the forcing procedure:
\begin{equation}
    \epsilon_\mrm{th} = \frac{4\tau_\mrm{f}\sigma_\mrm{f}^2N_\mrm{f}}{1+\tau_\mrm{f}k_0^{2/3}\epsilon_\mrm{th}^{1/3}} \label{eq:analyticalsolution}
\end{equation}
 $N_\mrm{f}=92$ represents the number of forcing modes and $\tau_\mrm{f}$ is chosen to be the Kolmogorov timescale $\tau_\mrm{\eta}$. Hence, the forcing amplitude $\sigma_\mrm{f}$ can be determined by inverting Equation \ref{eq:analyticalsolution} for the chosen values of $\Rel$ as:
\begin{equation}
    \sigma_\mrm{f} = \sqrt{\frac{\epsilon_\mrm{th} \cdot \left( 1+\tau_\mrm{f}k_0^{2/3}\epsilon_\mrm{th}^{1/3}\right) }{4\tau_\mrm{f}N_\mrm{f}}}
\end{equation}

For all the simulations carried out in this work, precursor simulations were perfomed to obtain a stabilized HIT field. As in Figure \ref{fig:epsilon}, the turbulent dissipation rate from the precursor simulations $\epsilon_\mrm{sim}$ are monitored over time, and the simulations are stopped when $\epsilon_\mrm{sim}$ sufficiently stabilizes to the analytical solution $\epsilon$ as in Equation \ref{eq:analyticalsolution}.
\begin{figure}[h]
    \centering
     \includegraphics[width=0.95\linewidth]{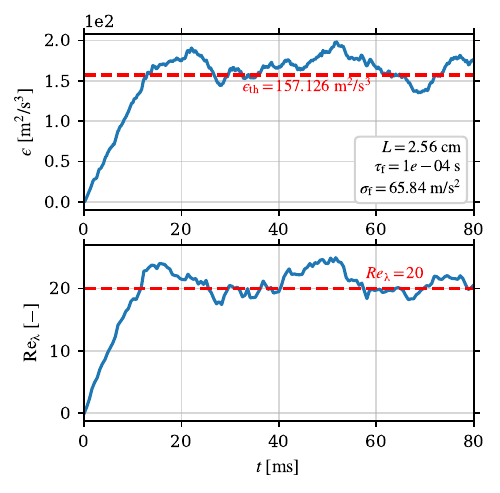}
    
    \vspace{-12pt}
    \caption{\red The evolution of turbulent dissipation rate $\epsilon$ (top) and Taylor Reynolds number $\Rel$ (bottom) over time in a cubical domain of size $L = \SI{2.56}{\centi\meter}$ containing air at $T_\mrm{g} = \SI{1200}{\kelvin}$ and $p=\SI{1.01325e5}{\Pa}$ with forced HIT corresponding to $\Rel = 20$ and $\eta=\SI{400}{\micro\meter}$. Annotated as a red line is the analytical value of the turbulent dissipation rate $\epsilon_\mathrm{th}$ as in Equation \ref{eq:analyticalsolution} and the assumed value of turbulent Reynolds number $\Rel$.}
    \label{fig:epsilon}
\end{figure}
}

\end{document}